\documentclass[colorlinks=true, linkcolor=blue, citecolor=blue, urlcolor=blue]{aa}
\usepackage{booktabs}
\usepackage{tabularx}
\usepackage{enumitem}
\usepackage{makecell}
\usepackage[switch]{lineno}
\usepackage[utf8]{inputenc}
\usepackage[utf8]{inputenc} 
\usepackage{graphicx}
\usepackage{multirow}
\usepackage{pifont}
\usepackage{txfonts}
\usepackage{natbib}
\usepackage{footnote}
\usepackage{longtable}
\usepackage{multirow}
\usepackage{booktabs}
\usepackage{xcolor}
\usepackage{changepage}
\usepackage{subcaption}
\usepackage{amssymb}
\usepackage{amsmath}
\usepackage{hyperref}
\usepackage{amsfonts}
\usepackage{arydshln} 
\usepackage{geometry}
\usepackage{longtable}
\usepackage{multirow}
\usepackage{booktabs}
\usepackage{pifont}
\usepackage{tabularx} 
\usepackage{natbib}
\usepackage{tablefootnote}
\usepackage{float}
\usepackage{acronym}
\usepackage{placeins}

\begin{document}

\title{X-ray Flares in Gamma-Ray Bursts at High and Very-High-Energies}

\titlerunning{MWL Study of flares in GRBs}

\author{Pawan Tiwari\thanks{\href{mailto:pawan.tiwari@gssi.it}{pawan.tiwari@gssi.it}} \inst{1,2},
Biswajit Banerjee \inst{1,2,3},  
Gor Oganesyan\inst{1,2,3},
Ansh Chopra \inst{1,2},
Annarita Ierardi\inst{1,2}, 
Marica Branchesi\inst{1,2,3}, 
Stefano Ascenzi \inst{1,2,4}, and
Maria Edvige Ravasio \inst{5,6,7} }

\institute{
\inst{1} Gran Sasso Science Institute (GSSI), Viale F. Crispi 7, L’Aquila (AQ), I-67100, Italy\\
\inst{2} INFN - Laboratori Nazionali del Gran Sasso, L’Aquila (AQ), I-67100, Italy\\
\inst{3} INAF - Osservatorio Astronomico d’Abruzzo, Via M. Maggini snc, I-64100 Teramo, Italy \\
\inst{4} INAF - Osservatorio Astronomico di Brera, via E. Bianchi 46, I23807 Merate (LC), Italy \\
\inst{5} Institute of Space Sciences (ICE), CSIC, Campus UAB, Carrer de Can Magrans s/n, Barcelona, E-08193, Spain \\
\inst{6} Institut d’Estudis Espacials de Catalunya (IEEC), Edifici RDIT, Campus UPC, Castelldefels (Barcelona), E-08860, Spain \\
\inst{7} Department of Astrophysics/IMAPP, Radboud University, Nijmegen, 6500 GL, The Netherlands\\}

\authorrunning{P. Tiwari et al.}   

\abstract
{A significant fraction of gamma-ray burst (GRB) early afterglows exhibit fast and bright X-ray flares, discovered by the X-ray Telescope onboard the Neil Gehrels Swift Observatory. Their rapid temporal variability and spectral evolution have suggested an internal shock origin, analogous to the prompt MeV emission, pointing to prolonged central engine activity. However, their physical origin and radiation mechanism remain debated. High-energy (> 100 MeV) and very-high-energy (> 30 GeV) gamma-ray observations provide a powerful probe of the flare dissipation region, constraining its size, magnetic field strength, and particle acceleration under the standard synchrotron self-Compton scenario in optically thin region of relativistic jets. In this work, we present a systematic multi-wavelength study of 66 X-ray flares from 47 GRBs observed by the Swift X-ray Telescope over 17 years, all occurring within the field of view of Fermi Large Area Telescope. We investigate their GeV counterparts and find that only five flares exhibit significant high-energy emission ($>3\sigma$). Broadband spectral modeling indicates that this GeV emission is consistent with the standard forward shock afterglow rather than associated with the X-ray flares. We further investigate correlations between the flare spectral properties and the energy fluxes at 1 keV, 10 keV, and 1 GeV. Using a synchrotron self-Compton model, we constrain the physical conditions of the emitting region, including the magnetic field strength, bulk Lorentz factor, and emission radius. For the most stringent GeV upper-limit cases, we find that the magnetic-to-electron luminosity ratio is $L_B/L_e \gtrsim 1$, implying highly magnetized emitting region. Finally, we predict the very-high-energy gamma-ray emission associated with X-ray flares occurring at both early ($\sim 500$ s) and late ($\sim 5000$ s) times after the GRB trigger and assess their detectability with Cherenkov Telescopes. We find that X-ray flares occurring at later times provide the most promising targets for follow-up observations, owing to the improved observational accessibility, sensitivity and reduced response-time constraints of facilities such as the Cherenkov Telescope Array Observatory.}

\keywords{high energy astrophysics, x-rays: flares, gamma rays: observations, methods: observational}

\maketitle
\section{Introduction} \label{sec:intro}
Gamma-ray bursts (GRBs) are one of the most energetic transient phenomena in the Universe, typically consisting of two emission phases: an initial highly variable prompt MeV emission followed by a multi-wavelength afterglow. In the standard internal–external shock scenario, the prompt MeV emission is produced by internal dissipation within a relativistic jet ~\citep{1994ApJ...430L..93R, 1997MNRAS.287..110S}, while the afterglow arises when the ejecta interact with the circumburst medium and generate a forward shock (FS) ~\citep{1993AAS...183.0417R, 1997ApJ...476..232M, 1998ApJ...497L..17S}. The external FS afterglow (hereafter afterglow) is expected to decay as a series of temporal power laws (PLs). However, thanks to detailed observations in the soft X-ray energy band with the X-Ray Telescope~\citep[XRT; 0.3-10 keV;][]{SWIFT:2005ngz} onboard the Neil Gehrels Swift Observatory~\citep[hereafter \textit{Swift},][]{2004ApJ...611.1005G}, the X-ray afterglow of GRBs has been found to exhibit several complex features. These include an early steep decay, extended plateau phases, and, most notably, X-ray flares (hereafter flares), which appear as sudden re-brightening in the light curve. Flares typically occur within the first few thousand seconds following the initial burst (prompt emission). They are found to occur in both long GRBs (lGRBs) and short GRBs (sGRBs). This makes flares a common feature in GRBs despite the difference in progenitors. 

Flares have been discovered in a significant fraction of GRBs~\citep{2007RSPTA.365.1213B, 2007ApJ...671.1903C, 2007ApJ...671.1921F}. These flares are typically narrow, with an average ratio of width to peak time of $\delta t /t \sim $   0.1~\citep{2007ApJ...671.1903C}. Typically, flare durations increase with time, while their intensities decrease~\citep{2018A&A...615A..80P}. Multiple flaring episodes can occur within a single GRB and are observed in a significant fraction of GRBs~\citep{2007ApJ...671.1903C, 2011ApJ...734L..27A}. 
 
Spectral studies using broad energy ranges and simultaneous data from XRT and \textit{Swift} Burst Alert Telescope~\citep[BAT; 15–150~keV;][]{Barthelmy:2005hs} indicate that flares cannot be described by a standard FS afterglow PL model. Instead, flare spectra often exhibit significant spectral evolution and are substantially harder than the underlying afterglow. In many cases, they cannot be adequately described by a single PL model, requiring  either a broken PL or an additional thermal component besides the non-thermal emission~\citep{2014ApJ...795..155P}. The inferred emission properties and outflow characteristics more closely resemble those of the prompt emission, supporting an internal origin for the flares rather than emission from the external FS~\citep{2007MNRAS.375L..46L, 2008PhDT........28M,2014ApJ...795..155P}. 

Despite detailed investigation and studies, a self-consistent model for the origin, energetics, and evolution of flares is still lacking. Proposed scenarios includes delayed central engine activity~\citep{2001ApJ...550..410M, 2005ApJ...630L.113K, 2006ApJ...642..354Z, 2006ApJ...636L..29P, 2006MNRAS.370L..61P, 2017MNRAS.464.4399D, 2018MNRAS.478.4323G}, an extended phase of the prompt emission~\citep{2016MNRAS.457L.108B}, late-time energy injection via refreshed shocks~\citep{1998ApJ...496L...1R},  and external shocks interacting with a dense or non-uniform circumburst medium~\citep{2017MNRAS.472L..94H}.

\begin{figure*}[ht]
    \centering
    \includegraphics[ width=\linewidth,
    trim={1.0cm 1.5cm 1.0cm 3cm},
    clip]{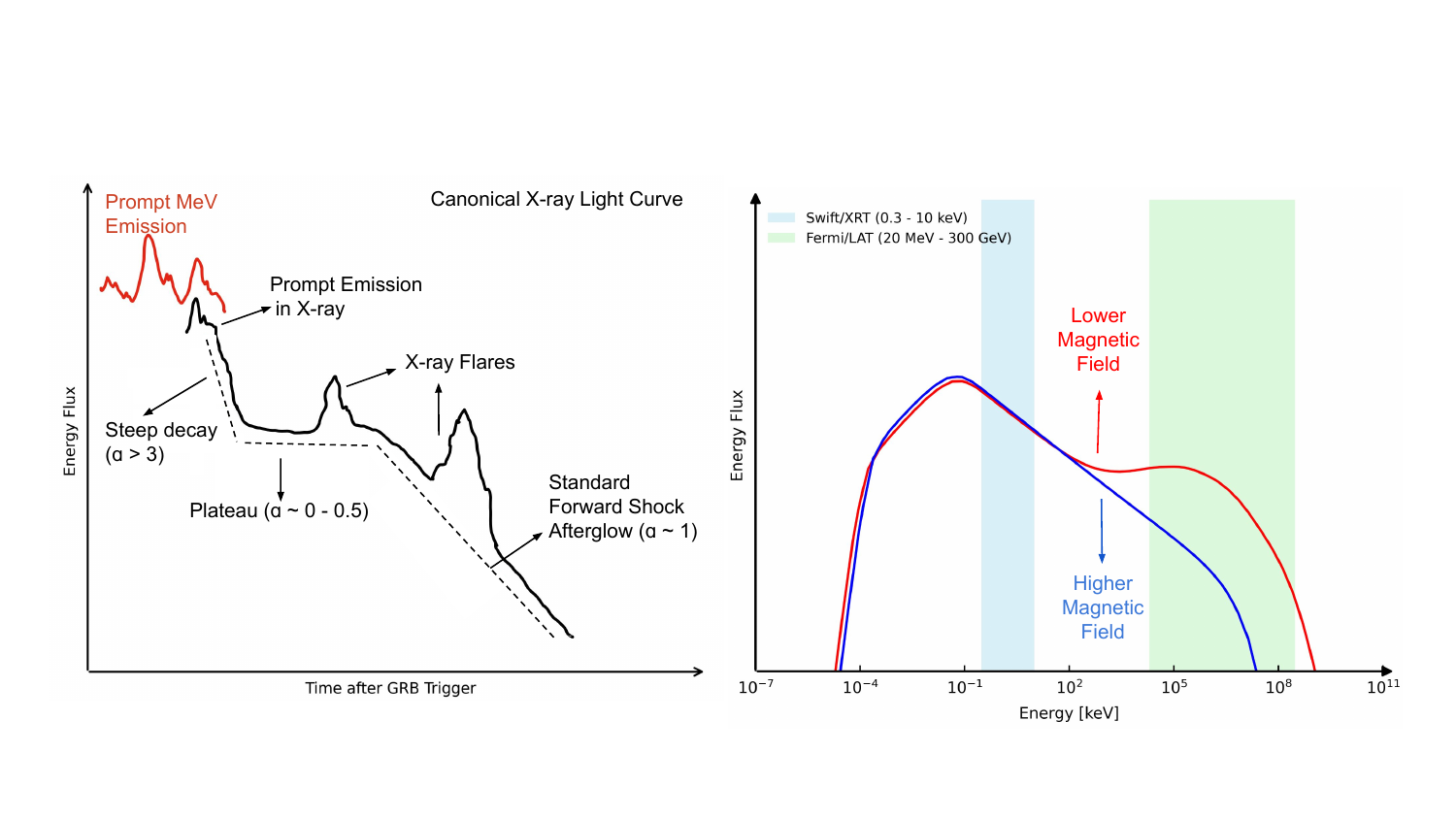}
    \caption{Illustration of the canonical X-ray light curve of GRBs with X-ray flares and the spectral energy distributions (SEDs) for different magnetic-field strengths. The left panel shows a schematic GRB X-ray light curve (with time and flux in arbitrary units), highlighting the prompt X-ray emission, steep decay  (($F_X \propto t^{-\alpha}$), with ($\alpha > 3$)), plateau phase ($\alpha \sim 0-0.5$), X-ray flares, and the standard forward-shock afterglow ($\alpha \sim 1$). The prompt MeV emission is shown in red. The right panel shows the predicted SEDs of an X-ray flare in the synchrotron self-Compton (SSC) scenario for two different magnetic-field strengths, while all other model parameters, including the flare duration, redshift, and bulk Lorentz factor, are kept fixed.}
    \label{fig:illustration}
\end{figure*}

The flares provide a natural probe of the dissipation region in relativistic outflows, where high-energy (HE; E $>$100 MeV) and very-high-energy (VHE; E $>$30 GeV) emission can be produced via standard synchrotron self-Compton (SSC) processes~\citep{2008A&A...480....5G}. In this scenario, synchrotron photons emitted by flare-accelerated electrons are upscattered to GeV–TeV energies by the same electron population. The resulting HE/VHE emission is highly sensitive to the physical conditions of the emitting region, in particular the magnetic-field strength (see Fig.~\ref{fig:illustration} for details). Alternatively, flare photons may also be inverse Compton (IC) scattered by relativistic electrons in the external FS, producing delayed HE emission in the external IC scenario \citep{2006MNRAS.370L..24F}.  

There have been very few extensive studies on flares in the context of GeV and MeV emission. To date, only one joint systematic study of X-ray and GeV emission for flares is that of~\citet{2015ApJ...803...10T}. The work is based on a relatively small sample covering only the first three years of observations after the launch of the Fermi Gamma-Ray Space Telescope~\citep[hereafter \textit{Fermi},][]{2007AAS...211.9803M}. Some individual bursts, such as GRB~100728A~\citep{2011ApJ...734L..27A, 2012ApJ...753..178H}, have shown flares accompanied by MeV–GeV emission detected by the Fermi Gamma-ray Burst Monitor~\citep[GBM; 8 keV - 40 MeV;][]{2009ApJ...702..791M} and Fermi Large Area Telescope~\citep[LAT; 30 MeV - 300 GeV;][]{2009ApJ...697.1071A}. The contemporaneous GeV emission has been interpreted as evidence for prolonged central-engine activity, with late internal shocks capable of producing both the X-ray flares and the HE component~\citep{2011ApJ...734L..27A,2013ApJ...772..152W}. However, an external FS origin for the GeV emission cannot be excluded, and IC scattering by electrons in the FS has also been proposed as a viable mechanism for the delayed HE emission~\citep{2011ApJ...734L..27A}.

In this work, we performed a systematic study of GeV emission originating from flares through a time-resolved spectral analysis including soft X-ray data from XRT and GeV data from LAT covering the period from 2008 to 2025. We compare the simultaneous energy fluxes~\footnote{In the rest of the paper, we denote the energy-flux by flux, when not mentioned otherwise.} (erg cm$^{-2}$ s$^{-1}$) in the X-ray and GeV energy ranges. We have also included hard X-ray data from the BAT (15–150~keV) when available in order to expand the spectra up to 150\,keV. 

To interpret the observed results, we employ the standard SSC framework in optically thin region of relativistic jets using the Leptonic Modeling Code~\citep[LeMoC;][]{Stathopoulos:2023qoy,2026arXiv260713242P}. We investigate the emission mechanisms responsible for producing X-ray and HE gamma-ray photons during flares. Although there is no direct observational evidence linking VHE photons to flares, their possible connection to prompt emission motivates us to predict the expected VHE emission and explore its detectability with current and future imaging atmospheric (or water) Cherenkov Telescopes. A recent study has modelled the multi-wavelength emission associated with flares, under the assumption of optically thin synchrotron radiation from non-thermal electrons in relativistic jets~\citep{2026arXiv260122749M}. In contrast, our work combines nearly two decades of X-ray--GeV observations of GRB flares with broadband SSC modeling to constrain the physical conditions of the emitting region and predict the associated VHE emission.

The paper is structured as follows: in Sect.~\ref{sec:sample_definition}, we describe the sample of GRBs selected for this study, together with the selection criteria. Section~\ref{sec:mwl_analysis} outlines the methodology adopted for the analysis of data from XRT, BAT, and LAT. The results obtained from our multi-wavelength analysis are presented in Sect.~\ref{sec:results}. An interpretation of these findings and prediction at VHE are provided in Sect.~\ref{sec:Intrepretation} and ~\ref{sec:prediction} respectively. Finally, Sect.~\ref{sec:Discussion} summarizes our main results and offers a discussion of their possible implications.

\section{Sample Definition} \label{sec:sample_definition}
The data sample used in this work consists of GRBs with detected X-ray flares observed between 11 June 2008 and 30 June 2025, spanning approximately 17 years. The initial sample comprises 332 GRBs containing a total of 478 X-ray flares. All flares were collected from the \textit{Swift} XRT online repository~\footnote{\url{https://www.swift.ac.uk/xrt_live_cat/}}, where they are reported based on the automated analysis pipeline~\citep{Willingale:2006zh}. We subsequently manually examined and cross-checked each flare to verify the reported flare activity.

Since the average \textit{Swift} XRT slewing time is approximately one minute, for GRBs with prompt emission lasting longer than $\sim$ 60 s, XRT can observe the prompt emission in the soft X-ray energy band (0.3 - 10 keV). These prompt emission pulses can be misidentified as flares (see Fig.~\ref{fig:illustration}). As the aim of this work is to study flares occurring during afterglow phase, we consider only flares detected at later times and excluded prompt emission pulses from our sample. The following selection criteria were applied in two steps to construct the final catalog:
\begin{itemize}
    \item We retained only flares occurring after the prompt-emission phase, defined using the \textit{Swift} BAT $T_{90}$ duration\footnote{The time interval between the epochs at which 5\% and 95\% of the total prompt-emission counts are detected.}. In addition, we excluded flares that were not completely covered by XRT observation (see Fig.~\ref{fig:incomplete_flare} for details). This is mainly due to the snapshot observing strategy of \textit{Swift}, Earth occultation, and observational gaps between consecutive XRT pointings~\citep{2007ApJ...671.1903C}. After applying these criteria, the sample consists of 206 GRBs with 276 flares, listed in Table~\ref{tab:flare_sample_after90}.
    \item For the selected flares, we searched for simultaneous \textit{Fermi} LAT observations. We retained only those flares satisfying the following criteria: (i) the source was within the \textit{Fermi} LAT field of view (FoV), (ii) the zenith angle~\footnote{The angle between an incoming photon's reconstructed direction and the local zenith line (a line originating at the center of the Earth and passing through the spacecraft)} was $<100^\circ$ , and (iii) the source off-axis angle~\footnote{The angular separation between the true/reconstructed direction of a source and the telescope's boresight.} was $<60^\circ$. 
\end{itemize}  
Applying these selection criteria results in a final sample of 47 GRBs containing 66 X-ray flares, of which 10 have measured redshifts ($z$). The final sample of flares with simultaneous \textit{Swift} XRT and \textit{Fermi} LAT observations is listed in Table~\ref{tab:lat_all}.

\section{Multi-wavelength Analysis} \label{sec:mwl_analysis}
\subsection{Fermi Gamma Ray Space Telescope}\label{sec:LATanalysis}
The Fermi Gamma-ray Space Telescope consists of two instruments: the GBM and the LAT, covering energies from about 8 keV to more than 300 GeV. The GBM is sensitive in the lower-energy range (8 keV – 40 MeV), while the LAT covers the HE gamma-rays from about ~30 MeV to > 300 GeV. LAT is a pair-conversion telescope comprising a 4$\times$4 array of silicon strip trackers and cesium iodide (CsI) calorimeters, shielded by a segmented anti-coincidence detector to suppress charged-particle background events. It has a FoV of $\sim$2.4 sr and scans the entire sky every 3 hours in survey mode~\citep{2009ApJ...697.1071A}.

For time-resolved spectral analysis, we used the \textit{gtburst} \footnote{\url{https://fermi.gsfc.nasa.gov/ssc/data/analysis/scitools/gtburst.html}} tool for extraction and processing of LAT data. We defined a $12^\circ$ region of interest (ROI) centered on each burst position and performed standard unbinned likelihood analysis for respective flare intervals. We used the \texttt{P8R3\_TRANSIENT020} event class, suitable for transient-source analysis, and the corresponding instrument response functions. As the spectral model, particle background, and Galactic component we assume \texttt{powerlaw2}, \texttt{isotr template}, and \texttt{template (fixed norm.)} respectively. We considered a minimum test statistic \footnote{Test statistic (TS) is  defined as -2 times the logarithm of the ratio of the likelihood for the model without the additional source (the null hypothesis) to the likelihood for the model with the additional source.} (TS\textsubscript{min}) of 10 which denotes the detection significance ($\sqrt{\rm TS}$ ~\footnote{\url{https://fermi.gsfc.nasa.gov/ssc/data/analysis/likelihood/ts_maps.html}}) of more than 3$\sigma$  . For each time bin in which the GRB was within the FoV of LAT, we calculated the 2$\sigma$ (95\% confidence level) upper limits (U.Ls.) for non-detections (TS$<$10) and the observed flux (F$\textsubscript{0.1-10~GeV}$) with 68\% errors (TS $\geq 10$). In addition, we also report flux at 1~GeV for comparison at specific energies. This value is derived from the PL fit performed over the 0.1–10 GeV energy range.

\subsection{Neil Gehrels Swift Observatory}
The Neil Gehrels Swift Observatory mission was launched in 2004 and has three instruments onboard that allow for multi-wavelength observations across hard and soft X-ray, ultraviolet, and optical wavebands: the BAT, the XRT and the UltraViolet Optical Telescope~\citep[UVOT;][]{Roming:2005hv}. The analysis techniques for XRT and BAT are described below. 

\subsubsection{\textit{Swift} XRT} \label{sec:XRTanalysis}
The XRT is a grazing-incidence-focusing X-ray telescope, covering an energy range of 0.3--10~keV. For the study, data are obtained from the Swift Science Data Center supported by the University of Leicester~\citep{Evans:2008wp}, include exposures taken in both Window Timing (WT) and Photon Counting (PC) modes.

For the analysis, we use the \texttt{XSPEC}~\citep{1996ASPC..101...17A} software package (v12.15.0), applying the C-statistic for parameter estimation. For all GRBs, we fit the spectra in the 0.3--10~keV band with a simple PL model, accounting for Galactic absorption using the multiplicative \texttt{tbabs} model. We adopt the Galactic neutral hydrogen column densities from the  GRB spectrum repository of XRT~\citep{2013MNRAS.431..394W}. For the time-resolved spectra, we also apply the \texttt{ztbabs} model to account for intrinsic absorption in the host galaxy, with the host galaxy hydrogen column density ($N_H(z)$) located at the burst redshift left as a free parameter. We report the best-fit parameters, unabsorbed flux (F$_{0.3-10~\mathrm{keV}}$), and photon index ($\Gamma_{\rm X}$) in Tab.~\ref{tab:xrt_all}. For GRBs with unknown redshift, we adopt $z=0$ when fitting the X-ray spectra~~\citep{Evans:2008wp}, and the best fit parameters are reported in Tab.~\ref{tab:xrt_all}. We also report flux at 1~keV and 10~keV for comparison. These values are derived from the PL fit performed over the 0.3–10 keV energy range.

\subsubsection{\textit{Swift} BAT} \label{sec:BATanalysis}
The BAT is a wide-field coded-mask telescope operating in the 15--150~keV energy range with a 1.4 sr field of view. The coded mask technique of BAT provides rapid positional accuracy of 1--4 arcminutes, typically within seconds of a GRB trigger.

We obtained hard X-ray data (15–150~keV) from the BAT archive, and processed them with the \texttt{batgrbproduct} pipeline\footnote{\url{https://www.swift.ac.uk/analysis/bat/}}. We generated the spectra using the \texttt{batbinevt} task and included systematic errors with the \texttt{batupdatephakw} and \texttt{batphasyserr} commands. We produced the response matrix with the \texttt{batdrmgen} task. Using \texttt{XSPEC}, we verified the exposure time of each spectrum to confirm data availability. We performed a time-resolved spectral analysis with \texttt{XSPEC} using $\chi^2$ statistics. We fitted the spectra using a PL model. 
The resulting best-fit parameters, fluxes, and photon indices ($\Gamma_{\rm B}$) are reported in Tab.~\ref{tab:batresult}. For non-detections, we report a flux U.Ls. with a 90~\% confidence level.

\begin{figure}[H]
    \centering
    \includegraphics[width=\linewidth]{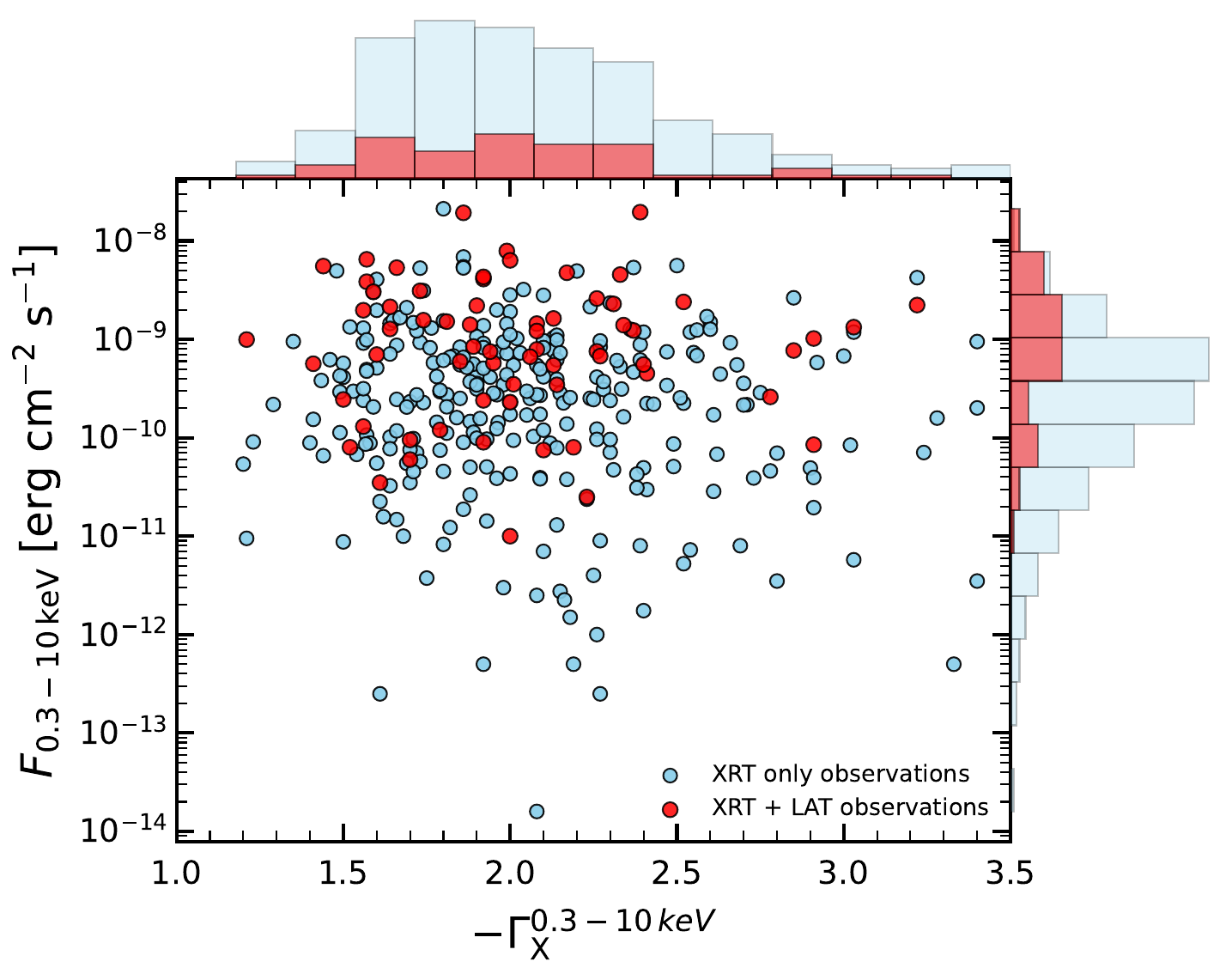}
    \caption{Comparison between the soft X-ray energy flux (F$_{0.3-10~\mathrm{keV}}$) and photon index ($\Gamma_{\rm X}^{0.3\text{-}10~\mathrm{keV}}$). The central panel shows the relation between F$_{0.3-10~\mathrm{keV}}$ and $\Gamma_{\rm X}^{0.3\text{-}10~\mathrm{keV}}$. Blue points represent GRBs with only \textit{Swift} XRT observations, while red points correspond to GRBs with simultaneous \textit{Swift} XRT and \textit{Fermi} LAT observations. The top and right panels show the corresponding distributions of $\Gamma_{\rm X}^{0.3\text{-}10~\mathrm{keV}}$ and F$_{0.3-10~\mathrm{keV}}$, respectively.}
    \label{fig:xrt_flux_index}
\end{figure}

\begin{figure*}[ht!]
    \centering
    \begin{subfigure}[b]{0.48\linewidth}
        \centering
        \includegraphics[width=\linewidth]{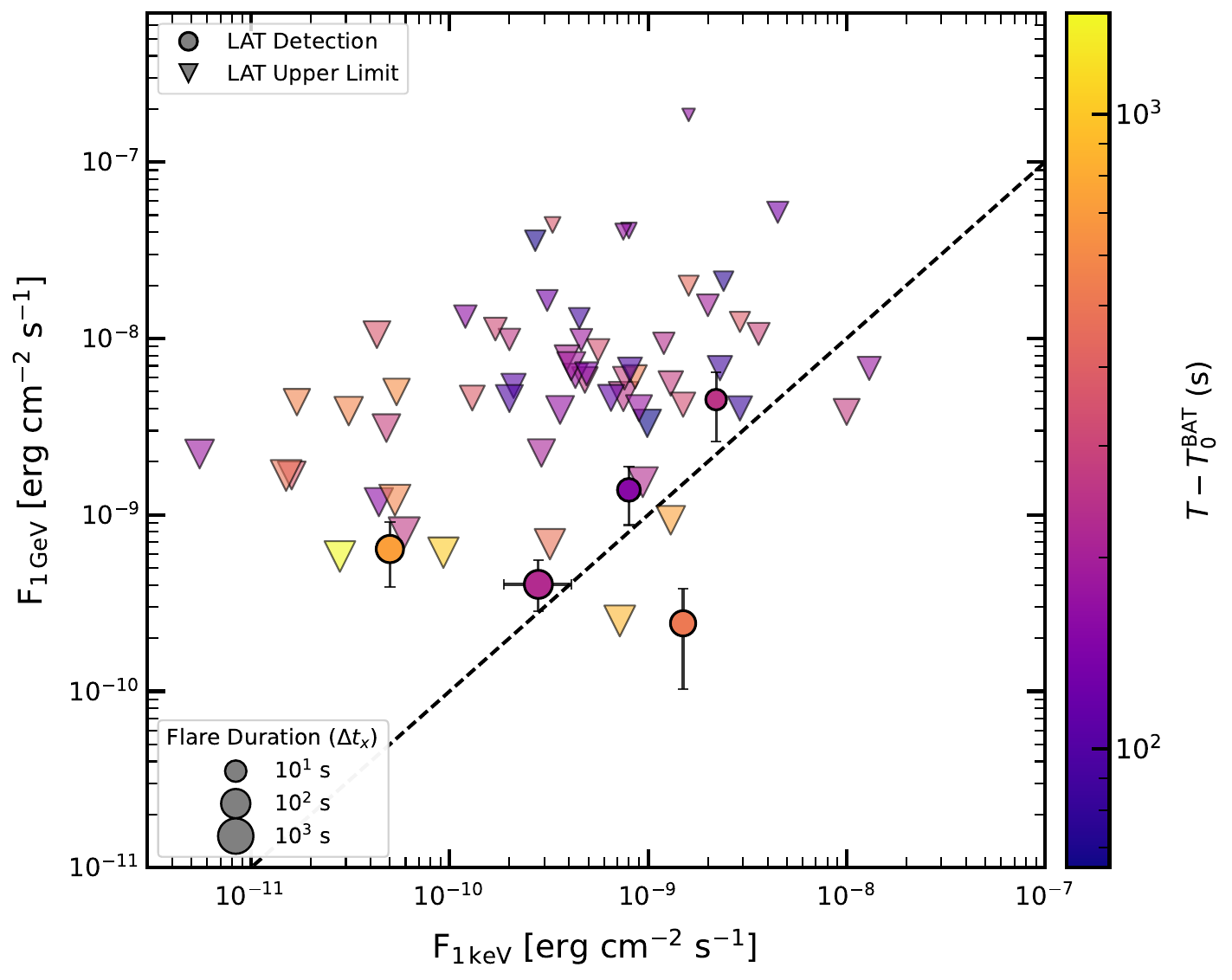}
        \label{fig:flux_1GeV_1keV}
    \end{subfigure}
    \hfill
    \begin{subfigure}[b]{0.48\linewidth}
        \centering
        \includegraphics[width=\linewidth]{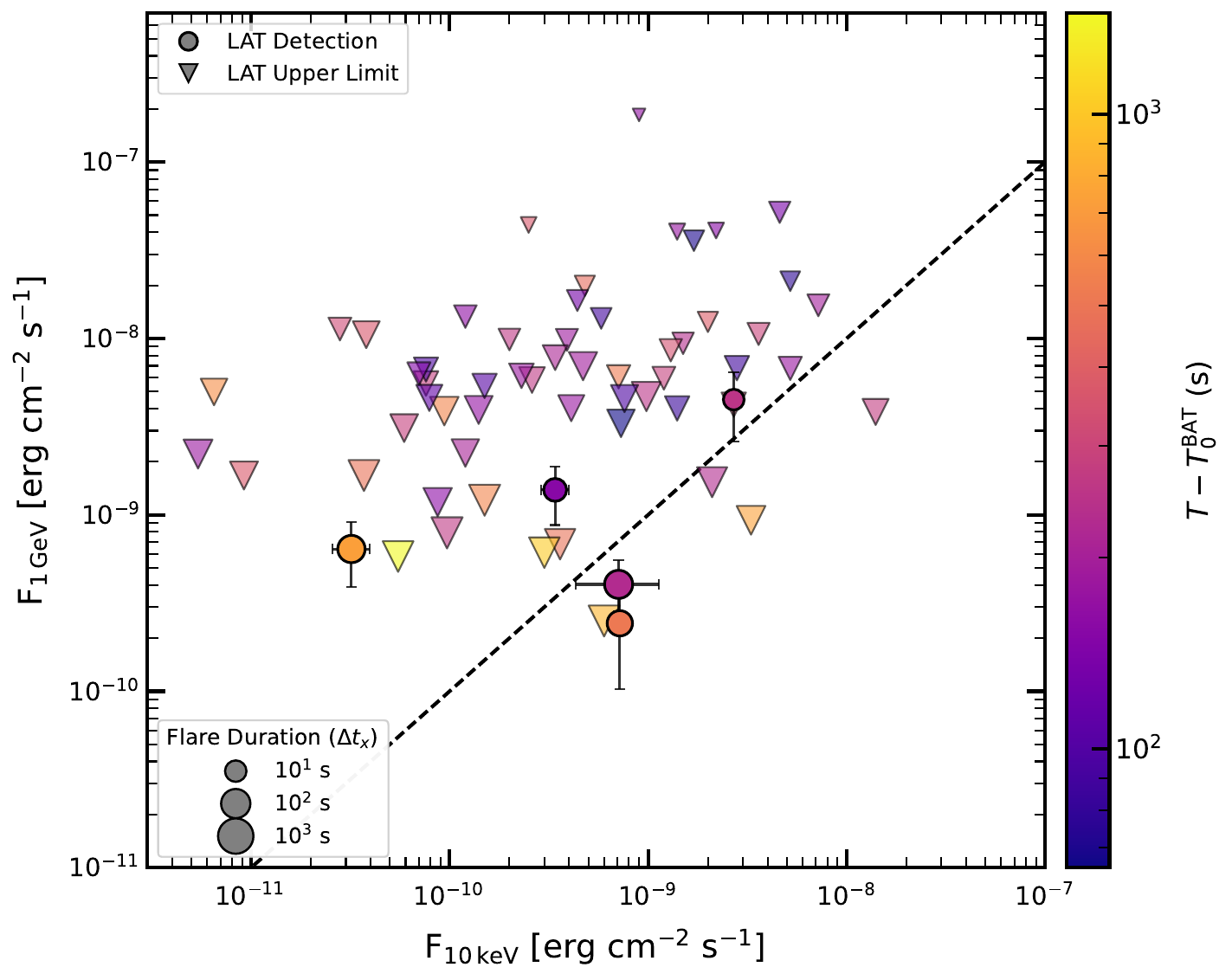}
        \label{fig:flux_1GeV_10keV}
    \end{subfigure}
    \caption{Comparison of HE gamma-rays with X-ray fluxes for flares.  The left panel shows the comparison of flux at 1 keV (F$_{1~\mathrm{keV}}$) with 1 GeV (F$_{1~\mathrm{GeV}}$). The right panel shows the comparison of flux at 10 keV (F$_{10~\mathrm{keV}}$) with 1 GeV (F$_{1~\mathrm{GeV}}$). LAT-detected flares are shown as points with error bars, whereas downward triangles represent the upper limits. The color gradient indicates the flare occurrence time after the GRB BAT trigger ($T-T_0^{BAT}$), and the marker size represents the flare duration ($\Delta t_x)$. The dashed black line denotes equality between the X-ray flux (1~keV in the left panel and 10~keV in the right panel) and the HE gamma-ray flux at 1~GeV.}
    \label{fig:flux_comparisonph_2_sidebyside}
\end{figure*}

\section{Results} \label{sec:results}
In order to characterize the broadband flare emission and extract relevant information on the properties of dissipation region, we used observable properties, such as fluxes and photon indices in X-ray and HE gamma-rays. In particular, we considered two representative energies for calculating flux in the X-ray band (1 and 10~keV) and one representative energy in the HE gamma-ray band (1~GeV). We also compared the flux and photon index distributions in the soft X-ray band (0.3--10~keV) between flares observed only with \textit{Swift} XRT and those with simultaneous \textit{Swift} XRT and \textit{Fermi} LAT observations.

\subsection{Energy Flux and Index Distributions in soft X-ray (0.3-10 keV)} \label{sec:xrt_distribution}
Using the XRT observations of 276 flares from 206 GRBs, we derived the unabsorbed energy fluxes (F$_{0.3-10~\mathrm{keV}}$), and photon indices ($\Gamma_{\rm X}^{0.3\text{-}10~\mathrm{keV}}$) in the 0.3–10 keV energy band. Fig.~\ref{fig:xrt_flux_index} shows the relation between the F$_{0.3-10~\mathrm{keV}}$ and - $\Gamma_{\rm X}$ for the GRB flare sample. The central panel presents the distribution of individual flares in the energy flux--photon index plane, while the top and right marginal panels show the corresponding one-dimensional distributions of the $\Gamma_{\rm X}^{0.3\text{-}10~\mathrm{keV}}$ and F$_{0.3-10~\mathrm{keV}}$, respectively.

The $\Gamma_{\rm X}^{0.3\text{-}10~\mathrm{keV}}$ span from approximately -1.5 to -2.5, indicating that the synchrotron peak energy ($E_{peak}^{synchroton}$) varies significantly across the sample, ranging from the optical band ($\sim$eV) to the hard X-ray band ($\sim10$--$150$ keV). Moreover, F$_{0.3-10~\mathrm{keV}}$ extends over nearly four orders of magnitude, ranging from $\sim10^{-8}$ to $\sim10^{-12}$ erg cm$^{-2}$ s$^{-1}$, with a few flares exhibiting even lower fluxes. The F$_{0.3-10~\mathrm{keV}}$ distribution peaks in the $10^{-10}$--$10^{-9}$ erg cm$^{-2}$ s$^{-1}$ range, which lies within the optimal sensitivity range of the LAT~\citep{2009ApJ...697.1071A}, making these flares well suited for searches for associated HE gamma-ray emission with \textit{Fermi} LAT.

\subsection{Comparison between X-ray and HE gamma-ray flux} \label{sec:flux_flux}
Using the joint observations of X-ray flares from XRT and LAT, we investigate the fluxes over a broad energy range, covering both the X-ray band (F$_{0.3-10~\mathrm{keV}}$) and the HE gamma-ray band (F$_{0.1-10~\mathrm{GeV}}$). In this section, we compare the HE gamma-ray flux at 1~GeV (F$_{1~\mathrm{GeV}}$) with the X-ray fluxes at two representative energies, 1~keV (F$_{1~\mathrm{keV}}$) and 10~keV (F$_{10~\mathrm{keV}}$) to investigate the presence of any correlation between the X-ray and GeV emission (see Fig.~\ref{fig:flux_comparisonph_2_sidebyside} for details). Table~\ref{tab:lat_all} and~\ref{tab:xrt_all} summarize the results of the time-resolved spectral analyses of the LAT and XRT data, respectively. GRBs with X-ray flares are divided into sources with known and unknown redshifts, separated by a dashed line in each table. 
 
The left and right panels of Fig.~\ref{fig:flux_comparisonph_2_sidebyside} compare the energy flux at 1 keV (F$_{1~\mathrm{keV}}$) with 1 GeV (F$_{1~\mathrm{GeV}}$) and 10 keV (F$_{10~\mathrm{keV}}$) with 1 GeV (F$_{1~\mathrm{GeV}}$), respectively. However, the number of data points with significant LAT detections (TS $>10$, corresponding to $\sim3\sigma$) is limited to five. This small sample size does not allow for a statistically robust assessment of trends or correlations. In particular, all five GeV-detected events belong to bright GRB in our sample: GRB 100728A, GRB 160325A, GRB 170810A and GRB 171120A, with fluence $S_{\gamma,\,1-1000\,\mathrm{keV}}$ \footnote{Fluence ($S_{\gamma}$; 1-1000 keV) is reported from ~\url{https://www.mpe.mpg.de/~jcg/grbgen.html}.} of $(12.70 \pm 0.01)$, $(1.86 \pm 0.02)$, $(1.0 \pm 0.6)$, and $(1.61 \pm 0.01)\times10^{-5}\,\mathrm{erg\,cm^{-2}}$, respectively. We also not that X-ray flares occurring at later times after the GRB BAT trigger tend to have longer durations. This correlation between the flare duration and its occurrence time has also been reported by \citet{2018A&A...615A..80P}.
As a result, later flares are integrated over longer time intervals, improving the sensitivity of the LAT analysis and enabling more stringent U.Ls. on the 1 GeV energy flux.

\section{Interpretation} \label{sec:Intrepretation}
\subsection{Does the GeV Emission come from X-ray Flares or the External forward shock Afterglow?} \label{sec:afterglow_flare}
The results presented in Sect.~\ref{sec:results} show that only five out of the 66 analyzed flares exhibit GeV emission. The remaining 61 flares, despite occurring within the LAT FoV, have no significant GeV detection. This raises the question of whether the GeV emission observed in these five cases is associated with the X-ray flare itself or originates from the external FS afterglow. 

Several studies~\citep{2014MNRAS.443.3578N,2015MNRAS.454.1073B} have shown that the early afterglow ($ t >100$~s) GeV emission observed in GRBs is predominantly associated with emission from the external FS. In the standard relativistic  SSC scenario, the GeV radiation can originate either from the HE tail of the synchrotron emission or from the SSC emission produced by the highest-energy electrons accelerated in the external shock. These electrons cool efficiently on a timescale shorter than the dynamical (expansion) timescale of the shock.  

\begin{figure}[h]
    \centering
    \includegraphics[width=\linewidth]{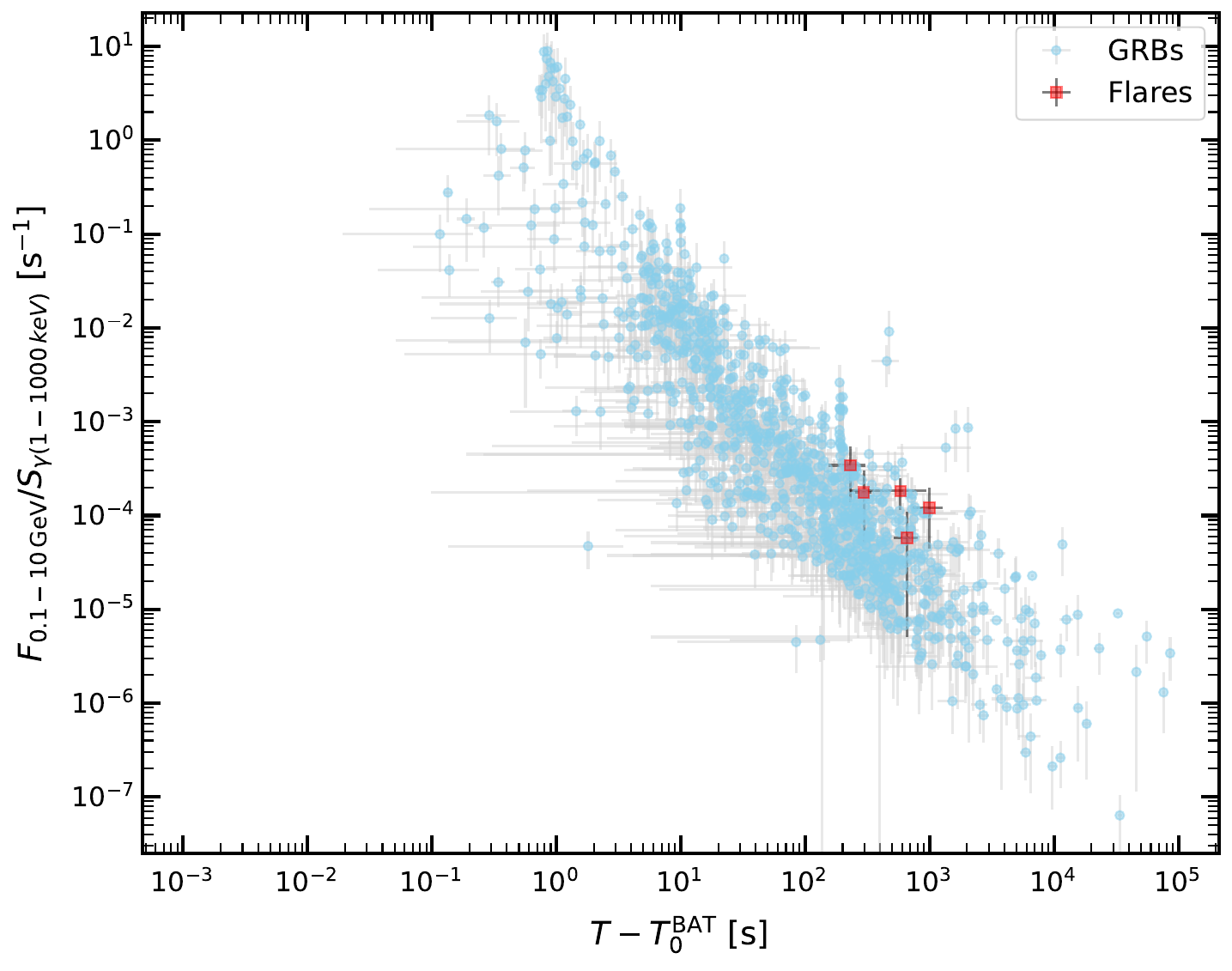}
    \caption{Ratio of the HE gamma-ray flux in the 0.1--10 GeV energy range, normalized to the prompt gamma-ray fluence (1--1000 keV), as a function of time since the GRB trigger. Circular blue markers represent HE gamma-ray measurements taken from \citet{2019ApJ...878...52A}, scaled to the 0.1--10 GeV energy band. Square red markers represent the flare candidates analyzed in this work. The flare sample includes two flare episodes from GRB~100728A and one flare episode each from GRB~160325A, GRB~170810A, and GRB~171120A.}
    \label{fig:flux_LAT_fluence}
\end{figure}

Several studies have shown that the normalized GeV afterglow light curves ($L_{\gamma}/E_{\rm iso}$) exhibit a common temporal evolution, indicative of clustering among GRBs~\citep{2014MNRAS.443.3578N,2019ApJ...878...52A,2025arXiv251005239T}. Using the observationally equivalent quantity ($F_{\gamma}/S_{\gamma}$), we tested the origin of the GeV emission detected during the flare epochs by examining the GeV flux normalized by the prompt-emission fluence as a function of time. The data points corresponding to the flare intervals (shown in red in Fig.~\ref{fig:flux_LAT_fluence}) are found to lie within the clustering region established for the GeV external FS afterglow population (shown in blue in Fig.~\ref{fig:flux_LAT_fluence}). This suggests that the GeV emission observed during the flare episodes is consistent with the expected FS afterglow component.

Therefore, the GeV emission detected during the flare intervals is likely dominated by the external FS afterglow emission, with any additional contribution from the flare component being subdominant. Consequently, in the following sections, we treat the measured GeV fluxes as U.Ls. on the HE emission associated with the flares and use them to perform broadband modeling and constrain the relevant microphysical parameters. These U.Ls. can provide constraints on the physical properties of the dissipation region, including the flare emission radius ($R_x$), bulk Lorentz factor ($\Gamma$), magnetic field strength ($B$), and electron luminosity ($L_e$), assuming that the observed emission originates from relativistic electrons.

In addition, we select GRB~100728A as a representative case study to further demonstrate that the observed GeV emission is dominated by the external FS afterglow component rather than by the flare emission itself (see Fig.~\ref{fig:flare_afterglow_100728A}) during flare epoch. This burst is chosen because it provides the most stringent GeV U.Ls. in our sample, allowing us to place the strongest constraints on the possible HE contribution from the flare component (see Tab.~\ref{tab:lat_all} and \ref{tab:xrt_all}).

\begin{figure*}[ht!]
    \centering
    \begin{subfigure}[b]{0.48\linewidth}
        \centering
        \includegraphics[width=\linewidth]{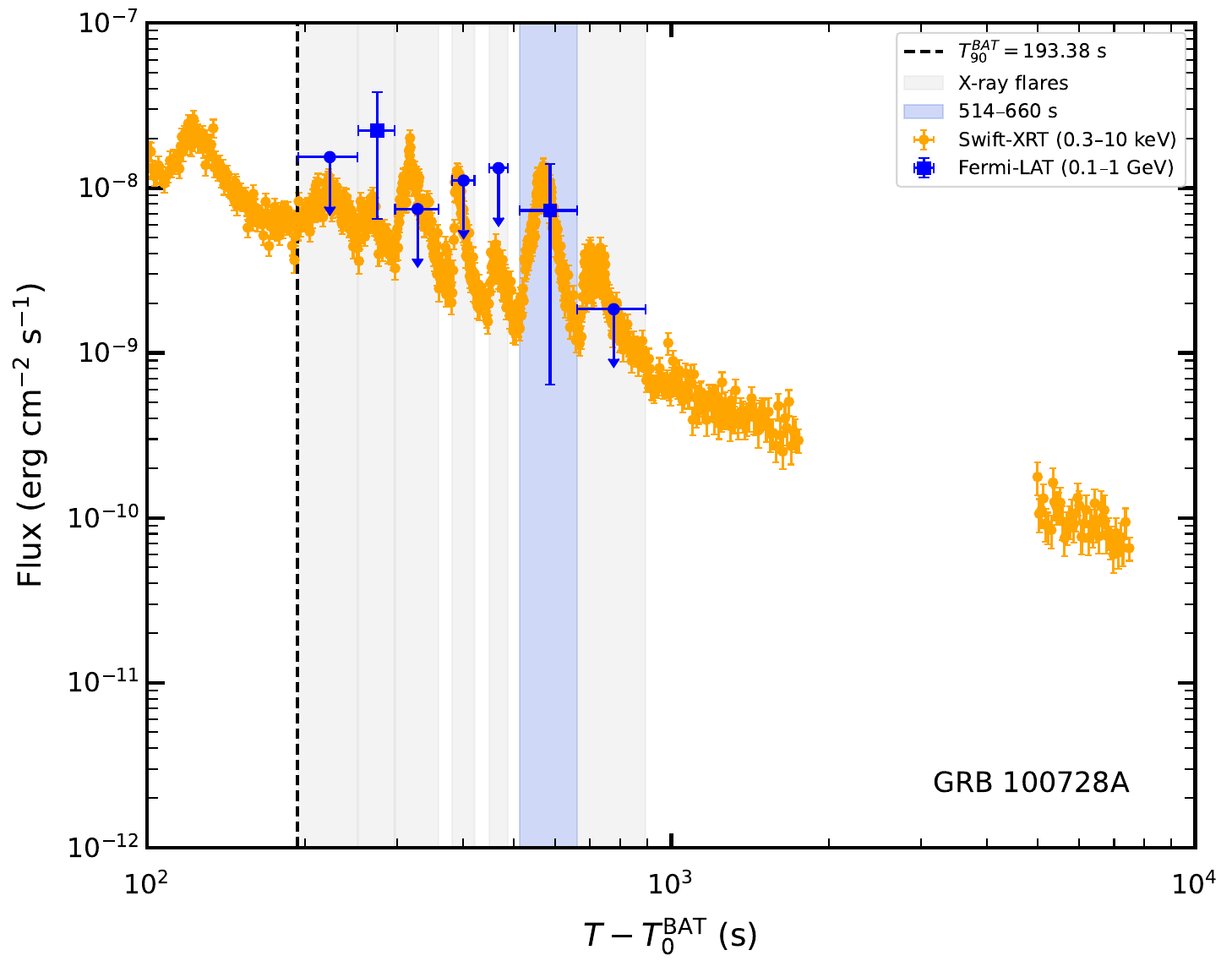}
        \label{fig:lc_100728A_model}
    \end{subfigure}
    \hfill
    \begin{subfigure}[b]{0.48\linewidth}
        \centering
        \includegraphics[width=\linewidth]{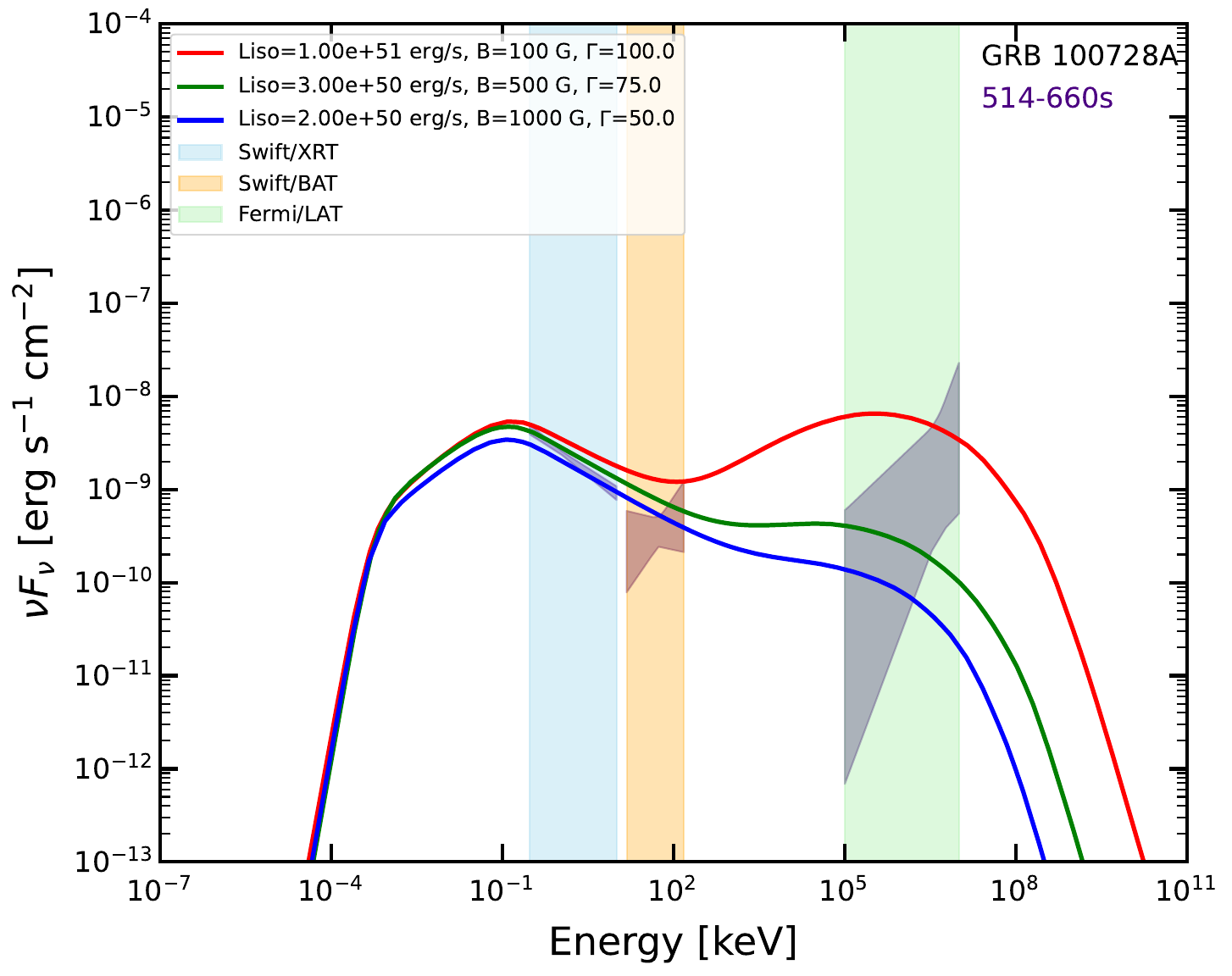}
        \label{fig:sed_100728A_model}
    \end{subfigure}
    \caption{\textbf{Left:} Multiwavelength light curve of GRB~100728A. The \textit{Swift} XRT (0.3--10 keV) light curve (orange points) is shown together with the \textit{Fermi} LAT (0.1--1 GeV) detections and upper limits during (blue points) the X-ray flare intervals. The shaded regions mark the X-ray flare intervals, while the blue-shaded region highlights the flare interval (514--660 s) selected for the spectral modeling presented in the right panel. The dashed vertical line denotes the \textit{Swift}-BAT $T_{90}$ duration. \textbf{Right:} Broadband spectral energy distribution of GRB~100728A during the selected flare interval (514--660 s) with SSC model fits for different values of the magnetic-field strength ($B$), bulk Lorentz factor ($\Gamma$), and electron luminosity ($L_e$). The model curves show that a broad range of physical parameters reproduces the observed X-ray flux while remaining consistent with the GeV upper limit. The shaded vertical bands indicate the energy ranges covered by \textit{Swift} XRT, \textit{Swift} BAT, and \textit{Fermi} LAT.}
    \label{fig:sed_100728A_model}
\end{figure*}

\subsection{Model Description} \label{sec:model_description}
With the GeV observations treated as U.Ls. on the flare emission, we proceed to investigate the microphysical parameters that govern the broadband emission from flares. We assume that the flares in X-rays originate from synchrotron radiation emitted by relativistic electrons in an optically thin region of the jet. Sub-photospheric dissipation models are not considered, as they are not expected to produce significant GeV emission because IC scattering is strongly suppressed in the Klein--Nishina regime~\citep{2011MNRAS.415.3693R}.

The flare emission is assumed to be produced in a shell of size $R_x$ and has a width $R_x/\Gamma$. The flare emitting region is assumed to contain shock-accelerated electrons injected with a PL energy distribution, $dN/d\gamma \propto \gamma^{-p}$ in comoving $B$, where $p$ is the electron spectral index and $\gamma$ is the electron Lorentz factor. The minimum and maximum Lorentz factors of the injected electrons are given by $\gamma_m \propto \left[\frac{\nu_m (1+z)}{\Gamma B}\right]^{1/2}$ where $\nu_ m$ is the characteristic synchrotron frequency corresponding to electrons with the $\gamma_m$ and
$\gamma_{max} \approx \left(\frac{6 \pi q_e}{\sigma_T B} \right)^{1/2}$, where $q_e$ is the electron charge and $\sigma_T$ is the Thomson scattering cross section.  Moreover, we consider the fast-cooling synchrotron scenario ($\nu_c < \nu_m$) as expected for flares produced in compact internal dissipation regions where the emitting electrons radiate their energy on timescales much shorter than the dynamical timescale~\citep{2005MNRAS.364L..42F, 2007RSPTA.365.1213B}. In this regime, the spectral energy distribution peaks at the characteristic synchrotron frequency corresponding to the minimum electron Lorentz factor, such that the synchrotron peak energy is $E_{\rm peak}=h\nu_m$.

The model is initially characterized by seven parameters: $R_x$, $\Gamma$, $B$, $\gamma_m$, $\gamma_{max}$, $p$ and $L_e$. We reduce the dimensionality of the parameter space using a set of physically motivated assumptions. First, we assume that the observed flare duration ($\Delta t_{\rm X}$) is given by the angular spread of emission i.e. $\Delta t_{\rm X} = R_x (1+z) / 2c\Gamma^2$, where c is the speed of light. This relation allows $R_x$ to be expressed solely in terms of the $\Delta t_{\rm X}$, $\Gamma$, and $z$. Second, for given $\nu_ m$, $\Gamma$, $B$ and $z$, the $\gamma_m$ is uniquely determined. The $\gamma_{max}$ is fixed by the Bohm's limit~\citep{Bohm1949, Meszaros2006, Lemoine2010}. Finally, we adopt a representative $p =2.5 $ and assume a high radiative efficiency, such that the injected electron luminosity is approximately equal to the observed flare luminosity ($L_e \simeq L_x$). With these assumptions, the number of free parameters is reduced to three: $B$, $\Gamma$ and $L_e$, for given $\nu_m$, $p$, $z$ and $\Delta t_{\rm X}$. In addition we defined the ratio between the Poynting flux luminosity ($L_B$) and $L_e$, as an important parameter characterizing the nature of the relativistic outflow. The Poynting-flux luminosity is given by $L_B = \frac{1}{2} R^2 \Gamma^2 c B^2$.

\begin{figure}[h]
    \centering
    \includegraphics[width=\linewidth]{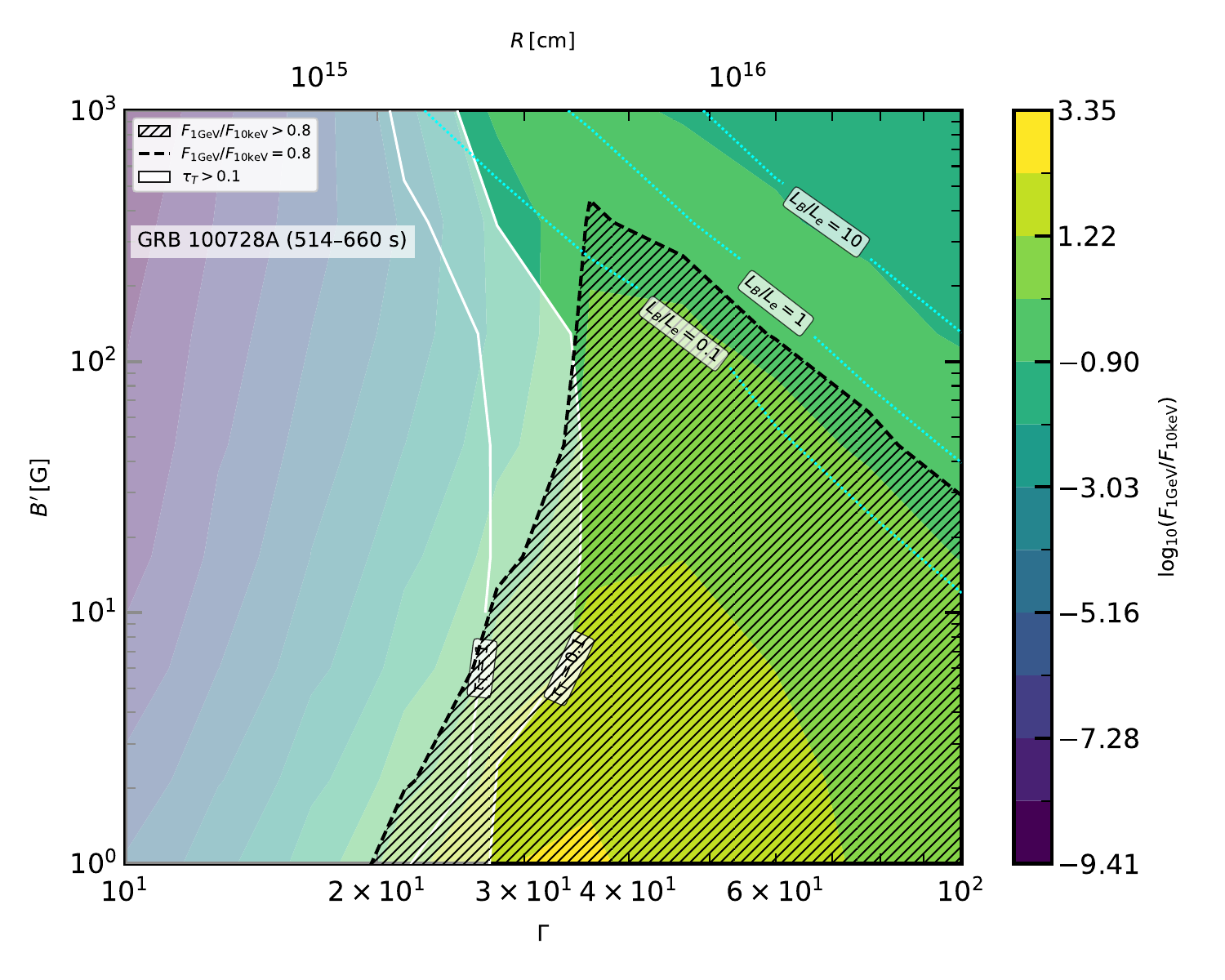}
    \caption{Allowed parameter space in the $\Gamma$--$B$ plane for GRB 100728A during the flare interval 514--660 s. Here $\Gamma$ is the bulk Lorentz factor and $B$ is the magnetic-field strength. The color map represents the ratio of HE gamma-ray to X-ray flux, while the shaded and bounded regions indicate constraints derived from the flare observations and SED modeling. The region enclosed by the black dashed contour indicates parameter values where the predicted 1 GeV-to-10 keV flux ratio exceeds or reaches the observed GeV upper limit. The translucent region represents the parameter space excluded by internal $\gamma\gamma$ pair production opacity. The remaining $(B,\Gamma)$ parameter space therefore represents the physically allowed solutions that are consistent with the observed X-ray emission, the GeV upper limit, and the requirement of high-energy photon transparency.}
    \label{fig:BG_100728A_model}
\end{figure}

We model the flare emission using the leptonic component of the Lepto-Hadronic Modeling Code~\citep[LeHaMoC;][]{Stathopoulos:2023qoy}. The code follows the evolution of relativistic electrons interacting with magnetic and radiation fields in an expanding spherical source, accounting for synchrotron emission, synchrotron self-absorption, IC scattering, photon-photon absorption, and adiabatic cooling. This framework enables the self-consistent calculation of broadband synchrotron and SSC spectra associated with flares in GRBs, connecting the observed emission to the underlying physical properties of the relativistic outflow.

\begin{figure*}[ht!]
    \centering
    \begin{subfigure}[b]{0.48\linewidth}
        \centering
        \includegraphics[width=\linewidth]{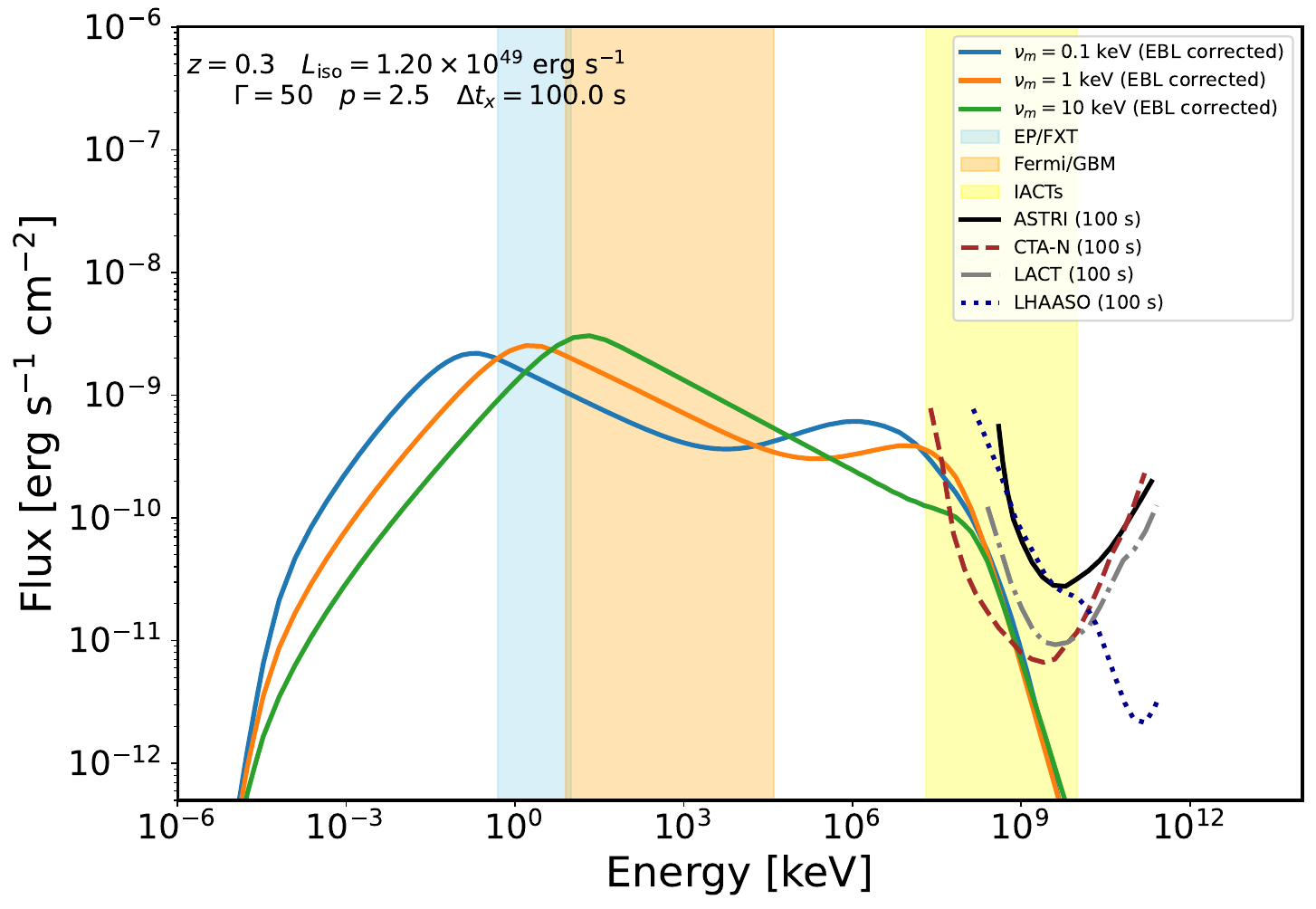}
    \end{subfigure}
    \hfill
    \begin{subfigure}[b]{0.48\linewidth}
        \centering
        \includegraphics[width=\linewidth]{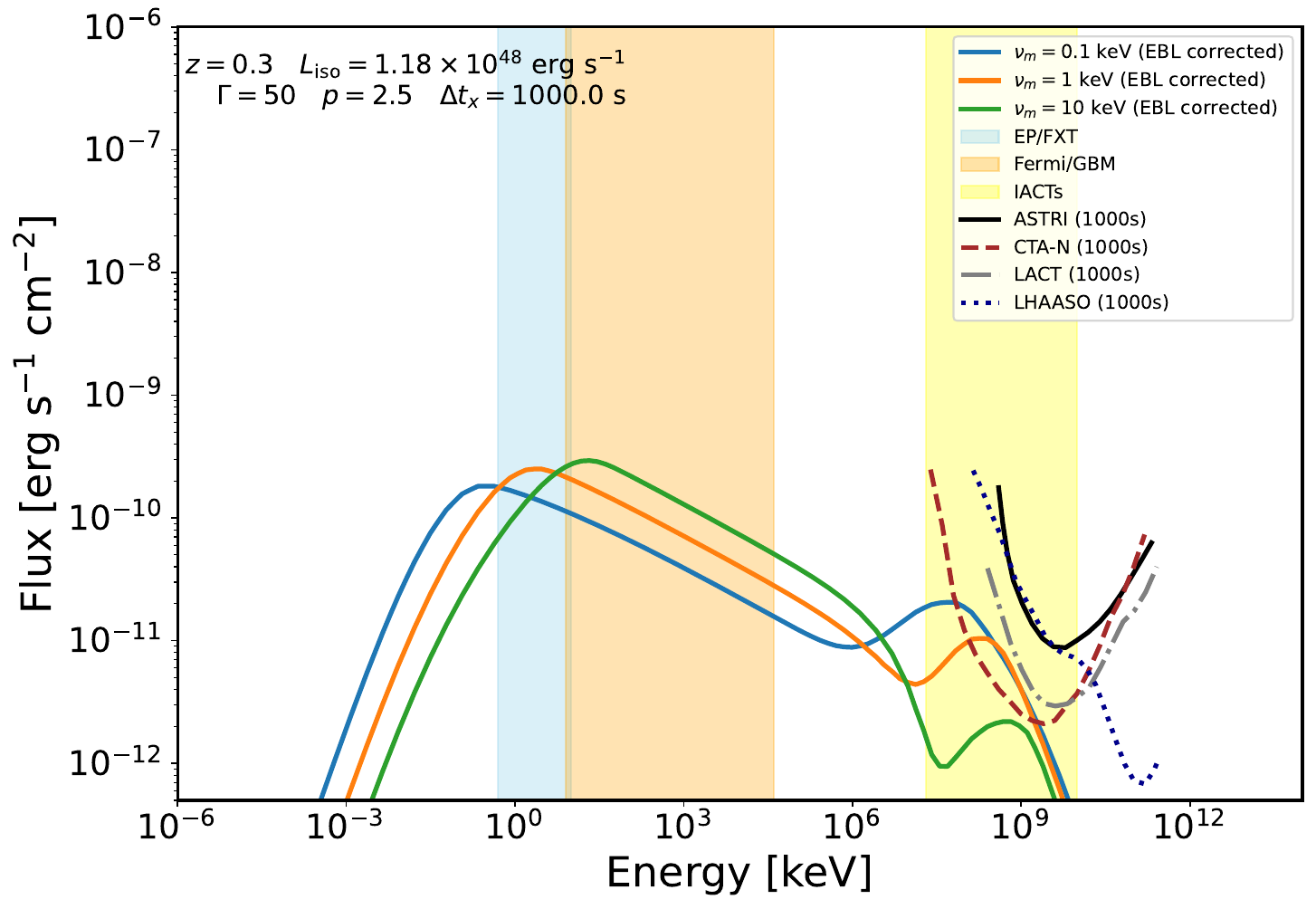}
    \end{subfigure}

    \caption{
    Spectral energy distributions (SEDs) associated with flares, computed for three different peak-energy scenarios with $\nu_{\rm m} = 0.1$, 1, and 10 keV, corresponding to different isotropic luminosities $L_{\mathrm{iso}}$. The model spectra are obtained using the SSC prompt-emission framework at redshift $z=0.3$, assuming fixed parameters $\Gamma = 50$ and electron spectral index $p=2.5$ with EBL correction. The dashed curves represent the sensitivity limits of different Cherenkov Telescopes. The left panel shows the sensitivity comparison for a 100 s exposure, while the right panel corresponds to a 1000 s exposure. These plots illustrate the detectability of the predicted HE/VHE emission from GRB flares for different observing times.}
    \label{fig:prediction_iacts}
\end{figure*}

\subsection{Microphysical parameters for flares} \label{sec:parameters}
To investigate the physical conditions of the dissipation region responsible for producing flares, we focus on the GRBs with measured redshifts and lie below equality line of Fig.~\ref{fig:flux_comparisonph_2_sidebyside}, as these provide the most stringent U.Ls. These bursts are: GRBs~100728A, 150821A, and 240912A. The flux comparison for the complete sample of GRBs with flares is presented in Fig.~\ref{fig:flux_comparisonph_2_sidebyside}. 

In this section, we focus on the most stringent U.L. case for a flare interval 514--660~s of GRB~100728A, for which XRT, BAT, and LAT data are available, allowing for detailed SED modeling. Fig.~\ref{fig:sed_100728A_model} (left panel) shows the multiwavelength light curve of GRB~100728A, including the identified flaring episodes. 

We note that reproducing the observed X-ray flux while remaining consistent with the GeV U.Ls. requires physically plausible combinations of the $L_e$, $B$, and $\Gamma$, as discussed in Sect.~\ref{sec:model_description}. Fig.~\ref{fig:sed_100728A_model} (right panel) presents the SED for the flare interval 514--660~s, together with the predictions of the SSC model for different combinations of $B$, $\Gamma$, and $L_e$. The model curves demonstrate that a wide range of physical parameters can successfully reproduce the observed X-ray emission while remaining below the GeV U.L. during the flare interval. We therefore explored the corresponding $(B,\Gamma)$ parameter space capable of explaining the observations.
We considered $B$ in the range $1$--$10^3$ G, $L_B/L_e = 0.1, 1, 10$,  and  $\Gamma$ in the range $10$--$100$. We also included the effects of internal $\gamma\gamma$ pair-production opacity. The generated electron–positron pairs increase the Thomson optical depth of the source, which is treated self-consistently in our model. By treating the observed GeV U.L as an upper bound on the flare-related HE emission, we constrained the allowed $(B,\Gamma)$ parameter space (see Fig.~\ref{fig:BG_100728A_model} for details).

In Fig.~\ref{fig:BG_100728A_model}, the region enclosed by the black dashed contour is excluded because it predicts a 1 GeV-to-10 keV flux ratio equal to or exceeding the observed U.L. The translucent region represents the region that is optically thick because of the pair-induced increase in opacity, as explained above. Consequently, the remaining $(B,\Gamma)$ parameter space represents the physically allowed solutions that are consistent with both the observed X-ray emission and the GeV U.L. We find that, for this flare (514-660s) of GRB~100728A which provides the most stringent GeV U.L constraint, the observed emission can be reproduced for the $L_B/L_e \gtrsim 1$. This implies that the flare emission region in jet is likely to be moderately to strongly magnetized. The $B$ and $\Gamma$ can vary depending on ratio of the $L_B/L_e$.

For the other cases with measured redshift lie below equality line of Fig.~\ref{fig:flux_comparisonph_2_sidebyside}, the light curve of GRBs with their flare, and the corresponding modeling results are presented in Appendix~\ref{sec:mp_other}.

\section{Prediction for VHE emission associated with flares} \label{sec:prediction}

The modeling of the flares emission component with SSC model in the broad energy range from X-rays to HE gamma-rays make flares particularly interesting to test a possible detection in the tens-hundreds GeV energy band by Cherenkov Telescopes. Unfortunately, none of the active ground-based instruments have reported the detection of the flares. Nevertheless, rather than the intrinsic absence of a VHE component, this could be easily explained considering the low duty cycle of IACTs, the large localization error of the event in the first minutes that prevent to perform a rapid follow-up~\citep{2026arXiv260407582M} and the occurrence of flares in GRBs. 

In this section, we predicted the broadband emission associated with flares, in particular VHE. The parameters we require are the $L_x$, $\Delta t_{\rm X}$, $B$, $\Gamma$, $\nu_m$,  and $L_B/L_e$. We adopt the empirical luminosity–duration relation reported by \citet{2016ApJS..224...20Y} to estimate the isotropic X-ray luminosity, $L_x$, as a function of $\Delta t_{\rm X}$. The relation is given by $\log L_x = A + B \log \Delta t_{\rm X}$, where $A = 51.12$, $B = -1.02$, and an intrinsic scatter of $\sigma = 0.83$ dex. Using this correlation, we predict $L_x$ at representative flare-width of $\Delta t_{\rm X} = 100$ and $1000$~s, corresponding to flares occurring approximately 500 s (early time flares) and 5000 s (late time flares) after the GRB trigger, respectively. 

\begin{figure*}[ht!]
    \centering
    \begin{subfigure}[b]{0.48\linewidth}
        \centering
        \includegraphics[width=\linewidth]{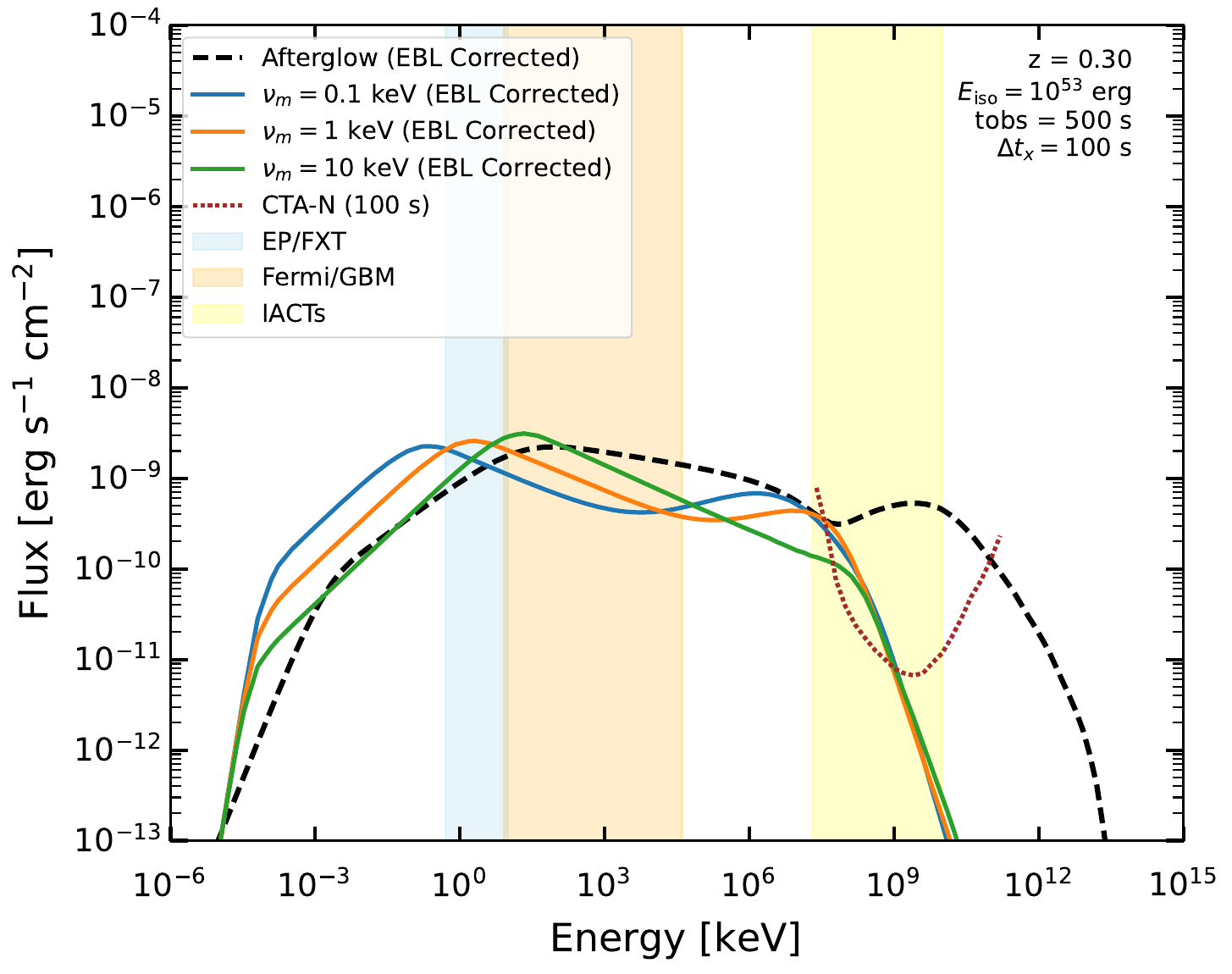}
    \end{subfigure}
    \hfill
    \begin{subfigure}[b]{0.48\linewidth}
        \centering
        \includegraphics[width=\linewidth]{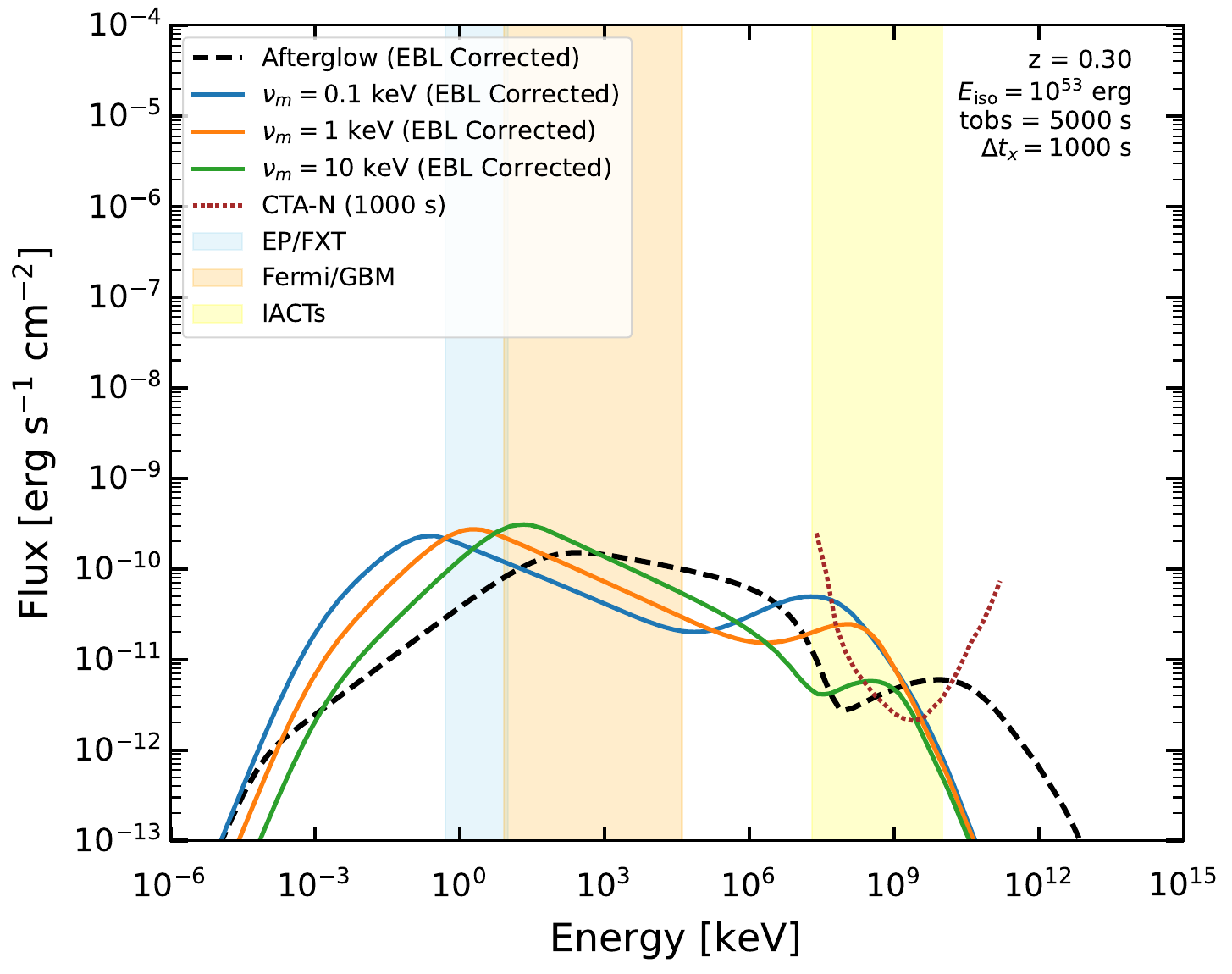}
    \end{subfigure}
    \caption{Comparison between the SEDs of the external forward-shock afterglow and X-ray flares at different epochs. The left panel shows the SED at an early epoch, assuming a flare variability timescale of 100 s, while the right panel shows the SED at a later epoch, assuming a variability timescale of 1000 s. The afterglow SED is calculated using typical GRB of $E_{iso} = 1\times 10^{53}$ erg and typical microphysical parameters $\eta = 0.1,\, \epsilon_{\rm B} = 10^{-4},\, \epsilon_{\rm e} = 0.1 ,p = 2.2$, and a wind-like circumburst medium ($A_\ast = 0.1$). The dashed curves (brown) represent the sensitivity limits of CTAO-N.}
    \label{fig:afterglow_flare}
\end{figure*}

For the bulk Lorentz factor, we adopt a representative value of $\Gamma$ = 50. We also assume a $L_B/L_e$ = 1, consistent with the results of this work (see Sect.~\ref{sec:parameters}). As shown in Fig.~\ref{fig:xrt_flux_index}, the $v_m$, can lie either below or above 1 keV. We therefore consider three representative values of the $v_m$: 0.1, 1, and 10 keV, to predict the SEDs of the flares. In addition, we adopt a representative $z = 0.3 $ to minimize the effects of attenuation by the extragalactic background light (EBL), using the EBL model of \citet{2011MNRAS.410.2556D}. The impact of EBL absorption on the observed emission is discussed in detail in Appendix~\ref{sec:EBL}. With these parameter choices, we estimate the intrinsic flare emission and the expected flux above 20 GeV.

We compared the expected VHE fluxes with the Cherenkov Telescope Array Observatory northern array differential sensitivities\footnote{Extracted from \url{https://www.cta-observatory.org/science/ctao-performance/\#1472563157332-1ef9e83d-426c}}(CTAO-N) assuming two different exposure times ($100$ s and $1000$ s, in Fig.~\ref{fig:prediction_iacts}). We also compared with other telescopes such as  ASTRI~\citep{2024Univ...10..146S}, LACT~\citep{2025arXiv250920694Z}, and LHAASO. All the sensitivities are obtained under standard assumptions, such as, dark sky condition, low zenith angle ($20^\circ$) and high atmospheric transmission.

We note that for the standard adopted set of input parameters found in this study for EBL corrected spectra, CTAO-N and LACT can play an important role in detecting the associated VHE emission from GRB flares, in particular flare occurring at later times ($\sim 5000$ s). It should be emphasized that, in most cases, the HE and VHE afterglow emission dominates over the flare emission (see Sect.~\ref{sec:afterglow_flare}). Therefore, while modeling the time intervals corresponding to the flare activity, one should simultaneously account for both the flare and afterglow contributions in order to obtain a physically consistent interpretation of the observed emission.

\subsection{Can we distinguish between X-ray Flares and External Forward shock Afterglows at VHE?}

As discussed in the previous section \ref{sec:afterglow_flare} and \ref{sec:parameters}, the GeV emission from the external FS afterglow generally dominates over that from the flare. Therefore, the modeling presented in the previous section becomes particularly relevant only if the VHE emission from flares exceeds the external FS afterglow contribution. To compare the flare and FS emissions, we calculated the external FS afterglow at early ($\sim 500 s$) and late ($\sim 5000 s$) epochs, corresponding to the flare occurrence times. We adopted the SSC model with the typical microphysical parameters inferred for TeV-detected GRBs, as listed in Tab.~\ref{tab:ssc_params_comparison}. We estimated SED for afterglow for typical GRB of $E_{iso} = 1\times 10^{53}$ erg and the typical microphysical parameters such as the index of the electron energy distribution $p$ = 2.2, the prompt efficiency $\eta$ = 0.1, fraction of shocked energy goes to electrons $\epsilon_{\rm e}$ = 0.1, the fraction of energy to the magnetic field $\epsilon_{\rm B}$ = $10^{-4}$. We consider a wind-like external medium with a density profile $\rho(R)\propto A_\ast R^{-2}$, adopting $A_\ast = 0.1$~\footnote{$A_\ast=1$ corresponds to a Wolf--Rayet stellar wind with a mass-loss rate of $\dot{M}=10^{-5}M_\odot\,{\rm yr}^{-1}$ and wind velocity $v_w=10^3\,{\rm km\,s^{-1}}$~\citep{1999ApJ...520L..29C}.}.

Fig.~\ref{fig:afterglow_flare} shows two cases with $\Delta t_x$ of 100~s and 1000~s, used to predict the broadband emission from flares under three different scenarios of $\nu_m$. The black dashed curve is the estimated FS afterglow SED. We find that when flare occurs at a relatively late epoch, approximately $\sim 5000$~s after the GRB trigger, with a typical duration of $\Delta t_{\rm X}\sim1000$~s, the flare emission can exceed the underlying afterglow at VHE (see the right panel of Fig.~\ref{fig:afterglow_flare}). In such cases, Cherenkov telescopes, particularly CTAO-N, offer the best prospects for detecting the flare emission. These late-time, bright X-ray flares therefore represent promising targets for future VHE observations, enabling direct tests of flare emission mechanisms and particle acceleration processes.

We also emphasize the importance of sensitive X-ray observatories in the coming decades for the systematic study of GRB flares. Current missions, such as the  Follow-up X-ray Telescope ~\citep[FXT, 0.3–10 keV;][]{2022hxga.book...86Y} onboard the Einstein Probe (EP) mission, provide enhanced X-ray monitoring capabilities that will enable the identification of faint and late-time flares. Coordinated multi-wavelength observations, combining X-ray, GeV, and VHE facilities, will be essential to fully understand the origin of flare emission and constrain the physical processes powering relativistic outflows.

\section{Discussion and summary} \label{sec:Discussion}
In this paper, we use multiwavelength observations from X-ray and GeV telescopes to investigate the emission processes during the flares occurring in afterglow phase of GRBs. These instruments collectively enable us to probe the emission over a broad energy range, spanning more than seven orders of magnitude from soft X-rays at 0.3~keV up to HE gamma rays above 100 GeV. Our analysis aims to probe the property of dissipation region for flares, including the magnetic field strength, in order to simultaneously explain the observed X-ray and GeV emissions. Additionally, we predict VHE emission associated to flares and their detectability with current and future Cherenkov Telescopes.

We selected a sample of 206 X-ray flaring GRBs with 276 flares observed during the period from June 2008 to June 2025. Since our primary goal is to search for GeV counterparts associated with X-ray flares, we further required that both the GRB and the corresponding flare intervals were within the field of view of the \textit{Fermi} LAT. Applying this criterion reduced the sample to 47 GRBs containing a total of 66 X-ray flares. 
 
Using the rich data sample described above, we performed an extensive time-resolved spectral analysis with simultaneous data in X-rays and HE gamma-rays. Using the XRT (0.3-10~keV) observations only, we find out that the photon-index distribution is concentrated mainly between -1.5 and -2.5, indicating that the synchrotron peak for flares can be from optical to hard X-rays. The corresponding flare fluxes in the 0.3-10 keV energy range span nearly four orders of magnitude, ranging from $\sim10^{-8}$ to $\sim10^{-12}$ erg cm$^{-2}$ s$^{-1}$, with a few events extending to even lower flux levels.
We compared the fluxes at F$_{1~\mathrm{keV}}$ with F$_{1~\mathrm{GeV}}$, and F$_{10~\mathrm{keV}}$ with F$_{1~\mathrm{GeV}}$ (see Fig.~\ref{fig:flux_comparisonph_2_sidebyside}). However, the number of data points with significant LAT detections (TS$>10$, corresponding to $\sim 3\sigma$) is limited to five. We find that the GeV emission observed during the flare intervals for these five events is broadly consistent with expectations from the external forward shock afterglow scenario. Thus, we conservatively treat the observed GeV flux as an upper limit on the flare contribution to constrain the microphysical parameters such as magnetic field strength of dissipation region, the emission radius, bulk Lorentz factor $\Gamma$, and the electron luminosity $L_e$. Using this assumption, we performed broadband SED modeling and derived the allowed parameter space for the $B$, $\Gamma$, $R_e$ and $L_B/L_e$ associated with the flare emission. An example of this analysis is presented for GRB~100728A (the deepest U.Ls case we have). We find that a wide range of parameter combinations can reproduce the observed flare spectra while remaining consistent with the GeV upper-limit constraints. Notably, we find that the $L_B/L_e$ is close to or above unity, suggesting an magnetically dominated jet. Moreover, we also performed modeling for GRBs with measured redshifts that lack significant GeV detections. However, in these cases, the relatively high LAT upper limits provide weaker constraints compared to the stringent U.L case.

Using the representative flare parameters derived in this work, we estimated the intrinsic VHE $\gamma$-ray emission expected from typical X-ray flares with $\Delta t_x $ of $\sim100$~s and $\sim1000$~s corresponding to early- and late-time flares occurring after the GRB trigger, respectively. The predicted SEDs indicate that nearby GRBs  with flares in our local Universe can produce detectable emission above $\sim20$ GeV. A comparison with the sensitivities of current and upcoming VHE facilities suggests that CTAO-N has promising capabilities for detecting such flare-associated VHE emission occurring at later epochs. However, since the HE/VHE afterglow component generally dominates over the flare contribution, a self-consistent modeling of both components is essential for interpreting the observed emission during flare intervals. A detailed treatment of the HE/VHE emission arising from the combined contributions of the afterglow and flare components is beyond the scope of the present work and will be investigated in a future study.

We summarize our work as follows:
\begin{itemize}[label=\textbullet, nosep, leftmargin=*]

\item We present a detailed time-resolved spectral analysis of X-ray flares in GRBs observed between 2008-2025 jointly observations in X-rays and GeV. With the help of a synchrotron self-Compton model, we attempted to explain the observables (energy fluxes and photon indices). 

\item From the observations, we find that the GeV emission detected during the flare intervals is generally dominated by the external forward shock afterglow component, with the flare-related emission likely embedded beneath it. Therefore, in order to properly interpret the observed high-energy emission during flare activity, both the afterglow and flare contributions must be simultaneously considered while modeling the HE/VHE emission.

\item The GeV upper limits place strong constraints on the physical conditions in the flare dissipation region. We find that a broad range of combinations of the magnetic field strength, bulk Lorentz factor, and emission radius can reproduce the observed X-ray flare properties while remaining consistent with the high-energy constraints. The inferred magnetic-to-electron luminosity ratio, $L_B/L_e$, is typically close to or greater than unity, favoring a magnetically dominated outflow. 

\item Our models predict that nearby GRBs hosting bright X-ray flares can produce detectable intrinsic VHE emission above $\sim20$ GeV. Among current and future facilities, CTAO-N is expected to provide the best prospects for detecting such emission, particularly for late-time flares. However, because the HE/VHE afterglow generally dominates the flare contribution, self-consistent modeling of both components is essential for robust interpretation of future observations. In synergy with CTAO and other next-generation VHE facilities, current and future sensitive X-ray observatories will be essential for systematically characterizing GRB X-ray flares and establishing their physical connection to HE and VHE gamma-ray emission. 

\end{itemize}

\begin{acknowledgements}
The authors acknowledge the use of public data from Fermi and Swift missions and thank the respective teams for making the data publicly available. PT acknowledges the Transnational Access (TA) Programme of the ACME project, which has received funding from the European Union’s Horizon Europe Research and Innovation programme under Grant Agreement No. 101131928. MB and BB acknowledge the same ACME project. This research acknowledges support from the ASI-GSSI contract n. 2025-5-U.0: “Gamma-ray bursts: a probe of multi-messenger and extreme Astrophysics”. MER is supported by the ”la Caixa” Foundation Junior Leader Fellowship (ID 100010434). The fellowship code is LCF/BQ/PI25/12100030. This work was also partly supported by the Spanish program Unidad de Excelencia Maria de Maeztu CEX2020-001058-M, financed by MCIN/AEI/10.13039/501100011033, and by the MaX-CSIC Excellence Award MaX4-SOMMA-ICE.
\end{acknowledgements}

\bibliographystyle{aa}
\bibliography{bibliography}

\clearpage
\appendix 
\section{Appendix} \label{sec:appendix}

This appendix provides supplementary material supporting the analysis and results presented in the main text. It includes additional figures, tables, and extended analyses that are not essential to the main text but are useful for completeness and reproducibility. We present the supporting material related to the flare sample selection, and statistical properties. This includes the full GRB sample used in this work, and the distribution of key observational quantities. We further provide results from the multiwavelength spectral analysis, including time-resolved XRT, BAT, and LAT analysis results for individual flares. 

\subsection{Selection of X-ray Flares}  \label{sec:selection_sample}
\begin{figure}[H]
    \centering
    \includegraphics[ width=\linewidth]{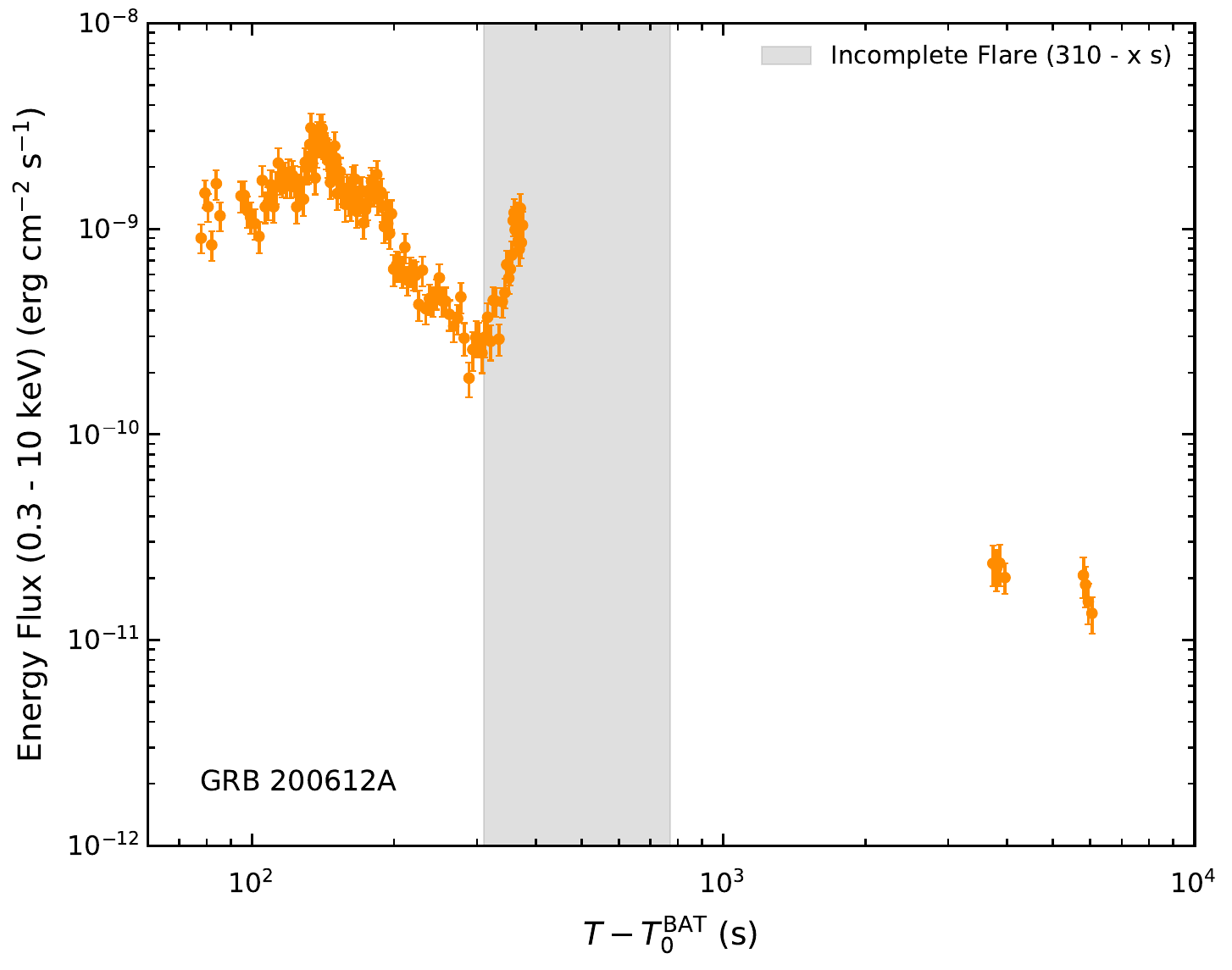}
\caption{GRB 200612A; An example of incomplete flares. The flare starts from 310s after trigger in BAT. These flares are excluded from this study.}
    \label{fig:incomplete_flare}
\end{figure}

In this section, we present the selection criteria adopted for the X-ray flares analyzed in this work. Table~\ref{tab:flare_sample_after90} lists all GRBs in our sample from 2008-2025, that exhibit flaring activity, as identified using the automated \textit{Swift} pipeline~\citep{Willingale:2006zh} and also manually examined. These flares occur after prompt emission phase ($T_{90}^{\mathrm{BAT}}$). For inclusion in our study, we apply the criterion that the flare evolution must be fully covered by \textit{Swift} XRT observations. Flares that are only partially observed or lack complete temporal coverage are excluded from the sample, as their peak fluxes and temporal properties cannot be reliably determined. An example of an excluded flare is shown in Fig.~\ref{fig:incomplete_flare}. 

\subsection{Multiwavelength Spectral Analysis Results}
In this section, we present the analysis results from the XRT, BAT, and LAT observations. Table~\ref{tab:lat_all} and ~\ref{tab:xrt_all} presents the results of time-resolved LAT and XRT spectral analysis for flares respectively. The flares intervals were chosen according to the criteria described in Section~\ref{sec:sample_definition}. Table~\ref{tab:batresult} presents BAT results for flares having BAT data. 

\subsection{GRB 100728A; SED Comparison between External Forward Shock Afterglow  with X-ray Flares}
\begin{figure}[H]
    \centering
    \includegraphics[width=\linewidth]{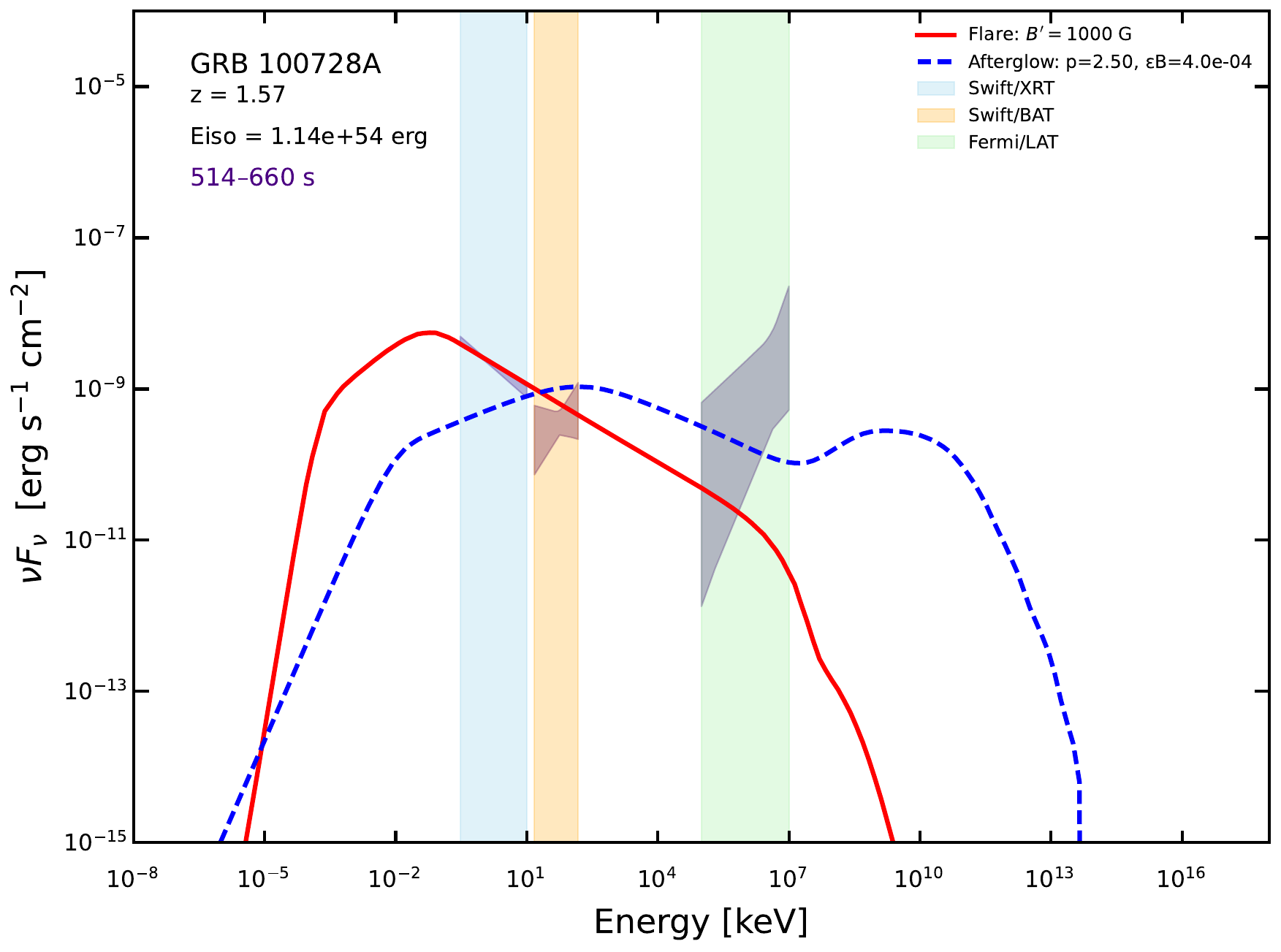}
    \caption{Comparison between the spectral energy distributions (SEDs) of the afterglow and flare emission for GRB~100728A. The afterglow component is modeled using $p = 2.5$, $\epsilon_{\rm B} = 4.0 \times 10^{-4}$, $z = 1.57$, and $E_{\rm iso} = 1.14 \times 10^{54}\,\mathrm{erg}$. The flare emission is calculated using the magnetic field strength $B$, $L_e$, and bulk Lorentz factor $\Gamma$ derived in this work.}
    \label{fig:flare_afterglow_100728A}
\end{figure}

As discussed in Sect.~\ref{sec:afterglow_flare}, the GeV emission observed in GRB~100728A for flare interval 514 - 660 s is likely dominated by the external forward-shock afterglow. Therefore, in this section we model the flare emission using the parameters derived in this work and compare it with the predicted forward-shock afterglow SED at the corresponding epoch. For the afterglow component, we adopt the parameters $p = 2.5$, $\epsilon_{\rm B} = 4.0 \times 10^{-4}$, $z = 1.57$, and $E_{\rm iso} = 1.14 \times 10^{54}\,\mathrm{erg}$~\citep{2021MNRAS.504..926M, 2025arXiv251005239T}.  We note that the flare component alone can successfully explain the observed X-ray emission, while the GeV emission during the flare epoch is better described by the combined contribution of the flare and the external FS afterglow, with the FS component dominating the HE emission (see Fig.~\ref{fig:flare_afterglow_100728A}). 

\subsection{Microphysical Parameters for Other Cases} \label{sec:mp_other}

\begin{figure*}[ht!]
    \centering
    \includegraphics[width=0.48\linewidth]{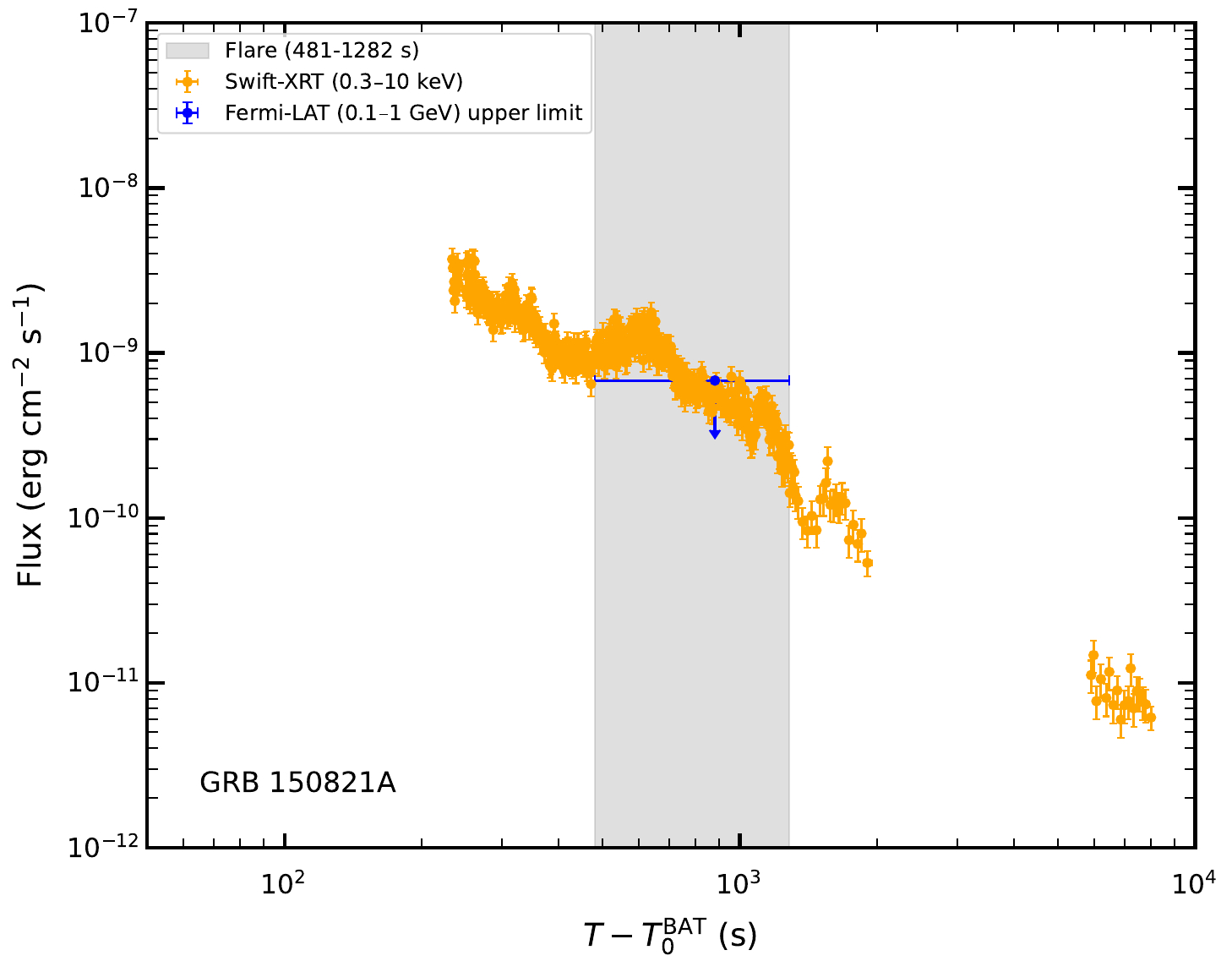}
    \hfill
    \includegraphics[width=0.48\linewidth]{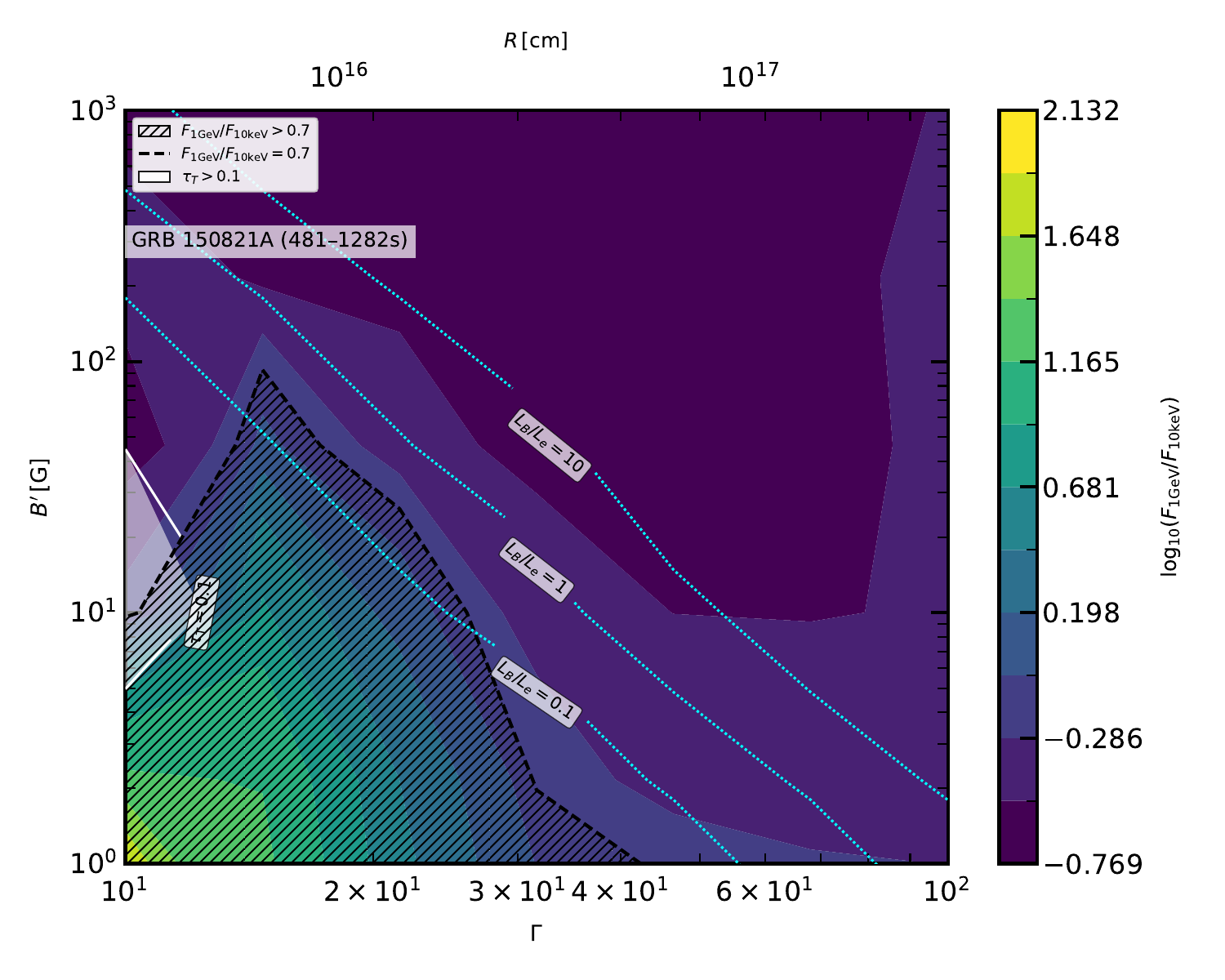}
    \caption{Left: X-ray light curve of GRB~150821A with Fermi LAT upper limit during flare intervals. Right: Constraints on $B$ and $\Gamma$ parameter space for GRB~150821A obtained from the observed X-ray flux and Fermi-LAT upper limit.}
    \label{fig:150821A_lc_bgamma}
\end{figure*}

\begin{figure*}[ht]
    \centering
    \includegraphics[width=0.48\linewidth]{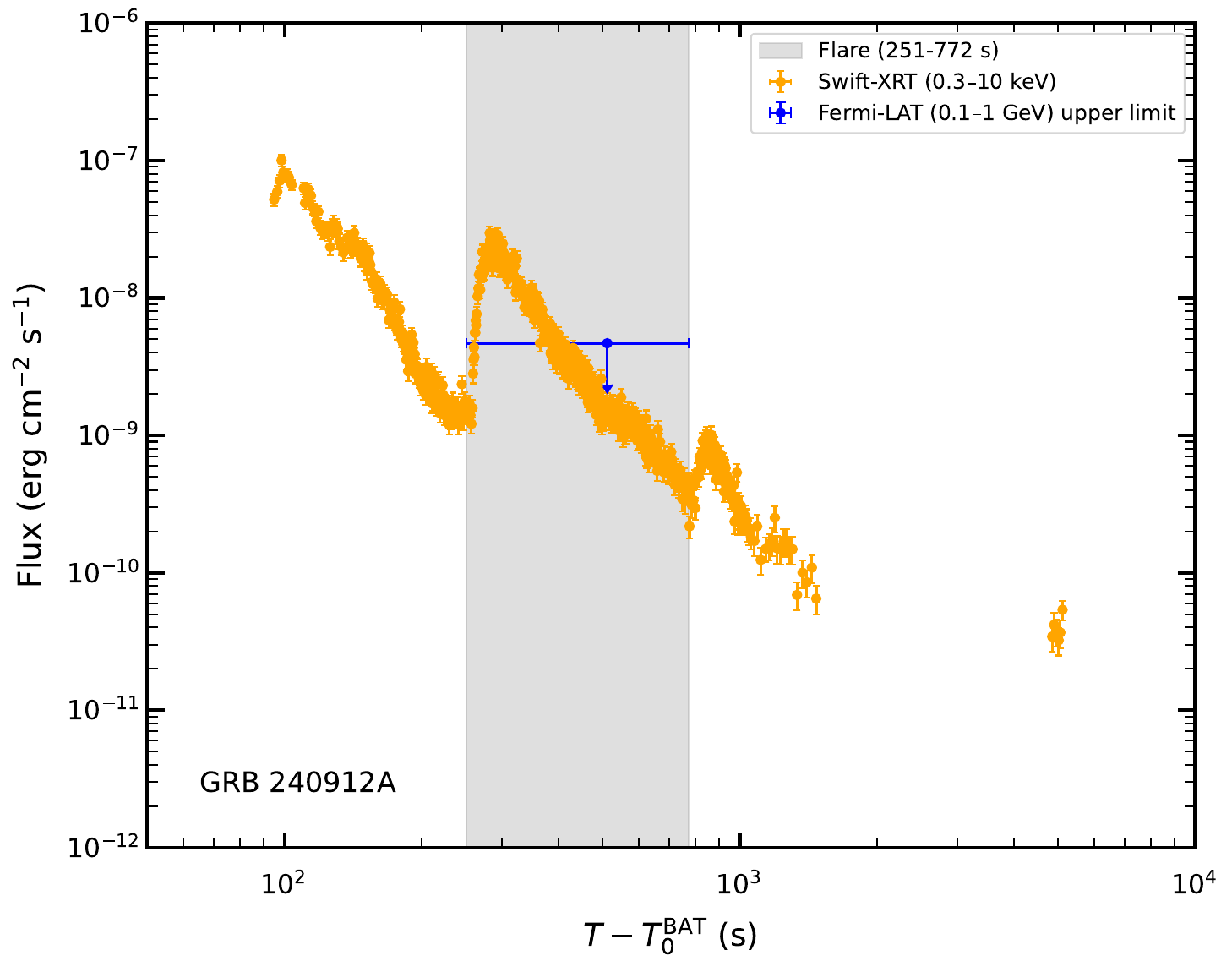}
    \hfill
    \includegraphics[width=0.48\linewidth]{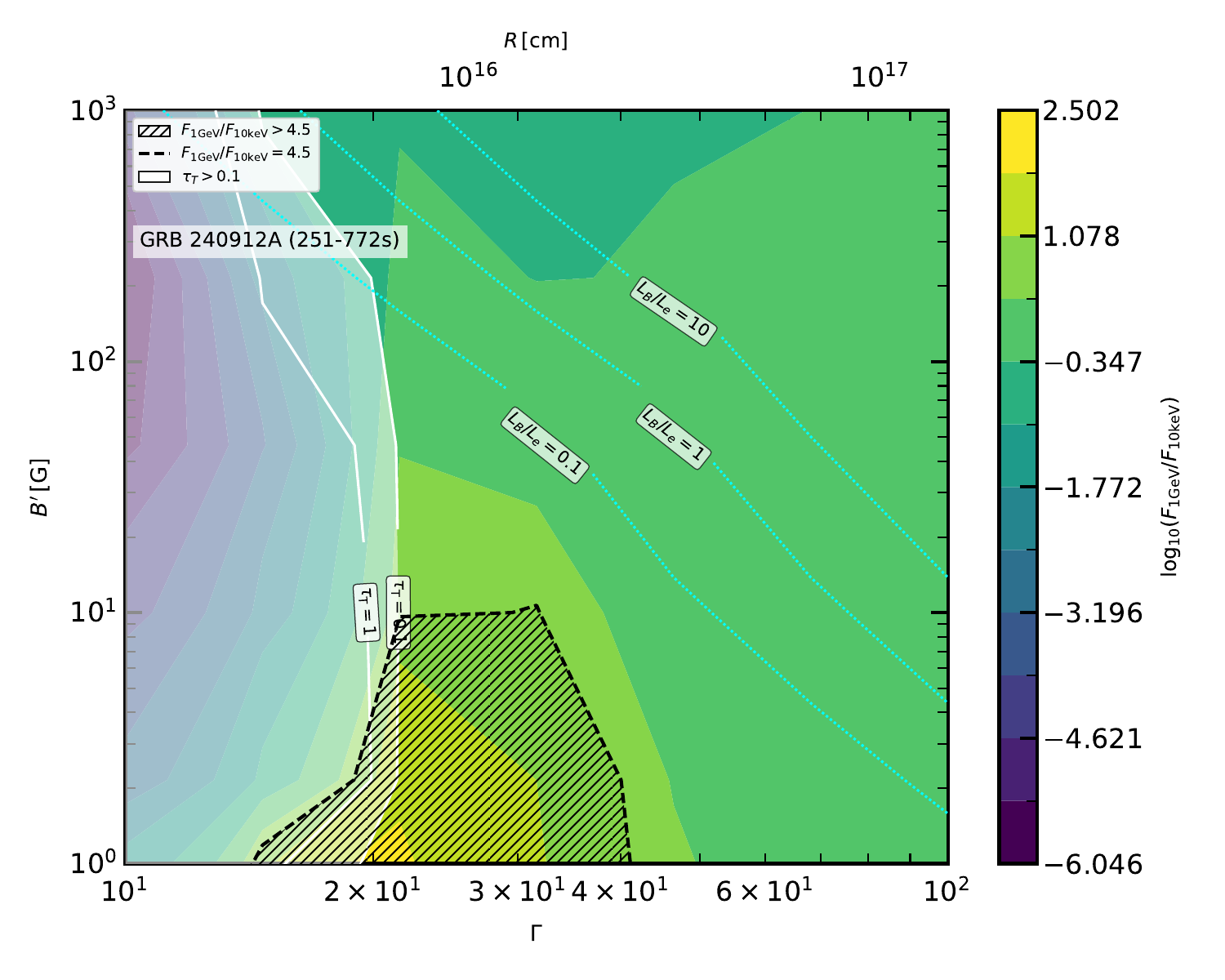}
    \caption{Left: X-ray light curve of GRB~240912A with Fermi LAT upper limit during flare intervals. Right: Constraints on $B$ and $\Gamma$ parameter space for GRB~240912A obtained from the observed X-ray flux and Fermi-LAT upper limit.}
    \label{fig:240912A_lc_bgamma}
\end{figure*}

As discussed in Sect.~\ref{sec:parameters}, we investigate the remaining GRBs that lie below the equality line in Fig.~\ref{fig:flux_comparisonph_2_sidebyside} by presenting their light curves and corresponding $B-\Gamma$ parameter spaces. These bursts include  GRB~150821A (see~Fig.\ref{fig:150821A_lc_bgamma}) and GRB~240912A (see~Fig.\ref{fig:240912A_lc_bgamma}).

\subsection{Effect of Extragalactic Background Light (EBL) on observed emission} \label{sec:EBL}

At VHE gamma-ray energies, the observed emission from GRBs or flares can be significantly affected by interactions with the EBL. During propagation from the event source to the observer, VHE photons can undergo photon--photon pair production with the low-energy photons of the EBL, resulting in an energy-dependent attenuation of the intrinsic gamma-ray spectrum. The strength of this absorption increases with both photon energy and source redshift, making the detection and interpretation of distant GRBs at VHE energies challenging.

The observed flux can be expressed as $F_{\rm obs}(E,z)=F_{\rm int}(E,z)\exp[-\tau(E,z)]$, where $F_{\rm int}$ is the intrinsic spectrum and $\tau(E,z)$ is the optical depth due to EBL absorption. Therefore, accurate modeling of the EBL attenuation is essential when interpreting VHE observations of GRBs and constraining their intrinsic emission properties.

\onecolumn

\scriptsize
\setlength{\tabcolsep}{6.7pt}
\begin{longtable}{l c c c c c p{4cm}}
\caption{Sample of Gamma-Ray Bursts with X-ray flares occurring after the prompt emission phase (for period 2008-2025). The horizontal dashed line separates GRBs with measured redshifts (above the line) from those without redshift measurements (below the line). The redshift values are taken from~\url{https://www.mpe.mpg.de/~jcg/grbgen.html}. \(T_{90}^{\text{BAT}}\) is the time interval between the detection of 5\% and 95\% of the total photons by BAT and reported from ~\url{https://swift.gsfc.nasa.gov/results/batgrbcat/}.}
\label{tab:flare_sample_after90} \\
\toprule
GRB name & BAT Trigger ID & Trigger Time (UTC) & $T_{90}^{BAT}$ (s) & $T_{90}^{BAT}$ err (s) & $z$ & Flare interval (s) \\
\midrule
\endfirsthead
\multicolumn{7}{c}{{\tablename\ \thetable{} -- continued from previous page}} \\
\toprule
GRB name & BAT Trigger ID & Trigger Time (UTC) & $T_{90}^{BAT}$ (s) & $T_{90}^{BAT}$ err (s) & $z$ & Flare interval (s) \\
\midrule
\endhead
\midrule \multicolumn{7}{r}{{Continued on next page}} \\
\endfoot
\endlastfoot
GRB240912A & 1253910 & 2024-09-12 01:46:39.185300 & 114.932 & 8.080113 & 1.234 & 134-184,251-772,800-1009  \\
GRB240529A & 1231488 & 2024-05-29 02:58:30.788060 & 158.972 & 12.97622 & 2.695 & 223-288  \\
GRB230414B & 1164180 & 2023-04-14 16:14:21.246980 & 29.06 & 7.740739 & 3.568 & 302-1092  \\
GRB220117A & 1093592 & 2022-01-17 16:18:51.546080 & 50.56 & 2.881111 & 4.961 & 158-254,486-4354  \\
GRB211211A & 1088940 & 2021-12-11 13:09:59.634260 & 50.712 & 0.9292406 & 0.0763 & 131-148,151-171  \\
GRB210504A & 1046782 & 2021-05-04 13:54:53.041040 & 142.984 & 9.329086 & 2.077 & 411-812  \\
GRB210420B & 1044382 & 2021-04-20 18:34:37.120380 & 158.16 & 29.6897 & 1.4 & 575-989  \\
GRB210411C & 1042398 & 2021-04-11 15:05:51.476880 & 12.8 & 0.5970595 & 2.826 & 141-414  \\
GRB201104B & 1004168 & 2020-11-04 17:32:54.779360 & 8.66 & 0.1618394 & 1.954 & 196-248  \\
GRB191221B & 945521 & 2019-12-21 20:39:13.151900 & 48.0 & 16.0 & 1.148 & 75.1-78.1,193-203,850-930,936-963  \\
GRB170531B & 755354 & 2017-05-31 22:02:09.205660 & 170.296 & 9.106015 & 2.366 & 540-880  \\
GRB170113A & 732526 & 2017-01-13 10:04:05.482200 & 20.296 & 4.332635 & 1.968 & 65.1-163  \\
GRB161219B & 727541 & 2016-12-19 18:48:39.308260 & 6.932 & 0.7931809 & 0.1475 & 92.3-121,305-720  \\
GRB160804A & 707231 & 2016-08-04 01:32:47.341920 & 152.744 & 19.20321 & 0.736 & 321-384,385-786  \\
GRB160227A & 676423 & 2016-02-27 19:32:08.096040 & 316.348 & 75.75925 & 2.38 & 372-1172  \\
GRB160117B & 670800 & 2016-01-17 13:59:27.649320 & 11.536 & 2.608028 & 0.87 & 48.2-70.4,75.4-178  \\
GRB151027A & 661775 & 2015-10-27 03:58:24.159080 & 129.58 & 5.465449 & 0.81 & 262-429  \\
GRB150915A & 655721 & 2015-09-15 21:18:24.860420 & 160.0 & 57.68882 & 1.968 & 161-215,216-235  \\
GRB150910A & 655097 & 2015-09-10 09:04:48.885620 & 112.272 & 36.48014 & 1.36 & 271-447  \\
GRB150821A & 652847 & 2015-08-21 09:44:00.976460 & 168.944 & 102.3371 & 0.755 & 233-286,288-364,481-1282  \\
GRB141109A & 618024 & 2014-11-09 05:49:55.236180 & 200.188 & 48.1482 & 2.993 & 218-241,304-400,448-498  \\
GRB140907A & 611933 & 2014-09-07 16:07:08.847220 & 80.0 & 35.77709 & 1.21 & 160-550  \\
GRB140518A & 599287 & 2014-05-18 09:17:46.630720 & 60.524 & 2.482379 & 4.707 & 74.8-97.3  \\
GRB140419A & 596426 & 2014-04-19 04:06:51.071220 & 80.076 & 3.775044 & 3.956 & 187-231  \\
GRB140304A & 590206 & 2014-03-04 13:22:31.098380 & 14.784 & 1.401371 & 5.28 & 81.3-109,276-599,747-960\\
GRB140301A & 589590 & 2014-03-01 15:24:49.478320 & 27.804 & 8.388573 & 1.416 & 81.0-131,514-834  \\
GRB140206A & 585834 & 2014-02-06 07:17:20.165420 & 94.188 & 18.36108 & 2.73 & 163-514  \\
GRB131117A & 577968 & 2013-11-17 00:34:04.717400 & 10.88 & 2.811633 & 4.042 & 82.1-128  \\
GRB131103A & 576562 & 2013-11-03 22:07:25.793680 & 15.208 & 3.034674 & 0.599 & 400-600,630-1000  \\
GRB131030A & 576238 & 2013-10-30 20:56:18.801700 & 39.424 & 3.674833 & 1.293 & 75.3-205  \\
GRB131004A & 573190 & 2013-10-04 21:41:03.688480 & 1.536 & 0.3263372 & 0.717 & 327-1002  \\
GRB130925A & 571830 & 2013-09-25 04:11:24.926960 & 160.296 & 3.390408 & 0.347 & 450-1179  \\
GRB130427B & 554635 & 2013-04-27 13:20:41.798740 & 25.904 & 5.971843 & 2.78 & 115-164  \\
GRB121229A & 544347 & 2012-12-29 05:00:21.961480 & 111.46 & 41.32595 & 2.707 & 386-619  \\
GRB121128A & 539866 & 2012-11-28 05:05:37.212960 & 23.428 & 1.649514 & 2.2 & 83.4-127  \\
GRB121027A & 536831 & 2012-10-27 07:32:29.746680 & 80.088 & 40.7569 & 1.773 & 141-194,219-520  \\
GRB121024A & 536580 & 2012-10-24 02:56:12.478460 & 67.968 & 31.72367 & 2.298 & 174-435  \\
GRB111228A & 510649 & 2011-12-28 15:44:43.604140 & 101.244 & 5.448429 & 0.714 & 136-140,150-176  \\
GRB111107A & 507185 & 2011-11-07 00:50:24.005480 & 31.072 & 7.152555 & 2.893 & 100-514  \\
GRB110205A & 444643 & 2011-02-05 02:02:41.367420 & 249.416 & 15.03472 & 2.22 & 355-364,465-489,596-684  \\
GRB101225A & 441015 & 2010-12-25 18:37:45.500280 & N/A & N/A & 0.847 & 4998-7447  \\
GRB100728A & 430151 & 2010-07-28 02:18:24.246260 & 193.384 & 10.57069 & 1.567 & 194-252,253-296,297-359,382-421,449-487,514-893  \\
GRB100621A & 425151 & 2010-06-21 03:03:32.840140 & 63.552 & 1.70755 & 0.542 & 134-250  \\
GRB100513A & 421814 & 2010-05-13 02:07:08.767460 & 83.496 & 21.62838 & 4.772 & 111-138,192-280  \\
GRB100425A & 420398 & 2010-04-25 02:50:45.077880 & 38.968 & 2.313897 & 1.755 & 72.5-77.5,111-160  \\
GRB100302A & 414592 & 2010-03-02 19:53:06.422100 & 17.948 & 1.644021 & 4.813 & 116-155,168-197,218-603  \\
GRB091029 & 374210 & 2009-10-29 03:53:22.595560 & 39.18 & 4.860979 & 2.752 & 261-685  \\
GRB090812 & 359711 & 2009-08-12 06:02:08.897520 & 74.496 & 15.28687 & 2.452 & 109-171,236-319  \\
GRB090709A & 356890 & 2009-07-09 07:38:34.592520 & 88.744 & 1.120514 & 1.8 & 248-314,355-468  \\
GRB090516A & 352190 & 2009-05-16 08:27:50.774420 & 181.008 & 41.25514 & 4.109 & 212-326  \\
GRB090423 & 350184 & 2009-04-23 07:55:19.349660 & 10.304 & 1.055547 & 8.2 & 157-403  \\
GRB090417B & 349450 & 2009-04-17 15:20:03.333200 & 266.94 & 35.38965 & 0.345 & 473-603,605-655,1229-5582  \\
GRB090404 & 348428 & 2009-04-04 15:56:30.440760 & 82.02 & 14.48431 & 3.0 & 107-144  \\
GRB081221 & 337889 & 2008-12-21 16:21:11.847580 & 33.908 & 0.9216507 & 2.26 & 93.8-159  \\
GRB081210 & 337073 & 2008-12-10 20:19:34.212340 & 145.908 & 7.028237 & 2.0631 & 284-477  \\
GRB081028A & 332851 & 2008-10-28 00:25:00.790760 & 284.424 & 30.54873 & 3.038 & 301-394,395-417,754-5542  \\
GRB081008 & 331093 & 2008-10-08 19:58:09.384500 & 187.824 & 38.22846 & 1.967 & 286-405  \\
GRB080928 & 326115 & 2008-09-28 15:01:32.866720 & 233.66 & 20.80139 & 1.69 & 339-405  \\
GRB080810 & 319584 & 2008-08-10 13:10:12.287980 & 107.668 & 3.471544 & 3.35 & 192-364  \\
\hdashline

\\[-0.2cm] 
GRB250516A & 1314210 & 2025-05-16 14:35:28.92 & 30.036 & 6.009782 & & 207-476 \\
GRB250207A & 1287821 & 2025-02-07 01:16:07.33 & 41.212 & 16.49658 & & 129-176,180-303 \\
GRB250108B & 1280056 & 2025-01-08 10:23:06.19 & 218.112 & 90.47834& & 492-978 \\
GRB240805B & 1246989 & 2024-08-05 14:28:54.12 & 64.996 & 2.550536 & & 71.3-77.9,129-173,272-513,1029-1806 \\
GRB240419B & 1223072 & 2024-04-19 12:22:52.33 & 144.552 & 12.22421& & 159-194 \\
GRB240101B & 1205744 & 2024-01-01 20:25:39.37 & 134.016 & 39.00793& & 214-631 \\
GRB231111A & 1195887 & 2023-11-11 14:17:19.71 & 34.076 & 4.85034  & & 167-400 \\
GRB231104A & 1194500 & 2023-11-04 01:47:39.91 & 46.712 & 1.568418 & & 594-973 \\
GRB230911C & 1191188 & 2023-09-11 08:47:25.51 & 14.488 & 2.858812 & & 684-813 \\
GRB230728A & 1181187 & 2023-07-28 02:50:03.22 & 14.76 & 0.8923072 & & 496-1500 \\
GRB230510A & 1167973 & 2023-05-10 12:06:28.58 & 178.368 & 6.931233& & 258-314 \\
GRB230405B & 1163119 & 2023-04-05 19:58:03.50 & 14.136 & 5.592688 & & 608-4838 \\
GRB230328B & 1162001 & 2023-03-28 14:54:48.54 & 22.336 & 5.710095 & & 94.4-259 \\
GRB230205A & 1152764 & 2023-02-05 10:29:51.18 & 2.3 & 0.3551057   & & 235-661 \\
GRB221202A & 1142997 & 2022-12-02 20:06:10.57 & 16.832 & 4.284177 & & 145-352,480-989 \\
GRB221024A & 1131029 & 2022-10-24 13:00:27.12 & 122.176 & 108.8623& & 142-426 \\
GRB220826A & 1121751 & 2022-08-26 11:55:21.74 & 11.14 & 1.019569  & & 512-730 \\
GRB220813A & 1120270 & 2022-08-13 19:23:03.57 & 16.592 & 1.912302 & & 131-285 \\
GRB220711B & 1115766 & 2022-07-11 18:16:28.69 & 87.056 & 9.321645 & & 126-519 \\
GRB220706A & 1114937 & 2022-07-06 16:09:14.29 & 85.996 & 18.3338  & & 536-2000 \\
GRB220618A & 1110821 & 2022-06-18 09:10:29.92 & 106.368 & 50.87108& & 185-245,282-405 \\
GRB220611A & 1110090 & 2022-06-11 18:01:51.17 & 57.0 & 12.36932   & & 152-445 \\
GRB220117B & 1093611 & 2022-01-17 20:05:28.15 & 24.368 & 4.633444 & & 165-295 \\
GRB210820A & 1069551 & 2021-08-20 17:38:00.70 & 151.74 & 20.5579  & & 177-189,193-209,221-469 \\
GRB210807C & 1064421 & 2021-08-07 22:57:08.49 & 89.972 & 6.255759 & & 197-386 \\
GRB210724A & 1061482 & 2021-07-24 20:14:09.24 & 44.176 & 5.377863 & & 175-412 \\
GRB210722A & 1061223 & 2021-07-22 20:54:42.02 & 50.204 & 10.53831 & & 78.1-222 \\
GRB210708A & 1059460 & 2021-07-08 02:25:37.08 & 2.772 & 0.7000486 & & 185-481 \\
GRB210419A & 1044032 & 2021-04-19 06:53:41.12 & 71.076 & 13.78015 & & 257-766 \\
GRB210410A & 1042113 & 2021-04-10 00:53:16.47 & 52.876 & 4.145965 & & 68.8-111 \\
GRB210402A & 1040123 & 2021-04-02 22:03:17.22 & 238.708 & 18.79081& & 648-3440 \\
GRB210305A & 1035922 & 2021-03-05 19:03:04.45 & 67.524 & 4.210722 & & 96.3-136 \\
GRB210209A & 1031636 & 2021-02-09 21:39:44.52 & 139.408 & 21.51339& & 202-4618 \\
GRB201029A & 1003002 & 2020-10-29 20:19:50.10 & 491.024 & 17.62968& & 1015-1859 \\
GRB200806A & 987016 & 2020-08-06 15:28:49.91 & 38.688 & 1.004064  & & 199-280 \\
GRB200729A & 984929 & 2020-07-29 19:38:05.09 & 122.0 & 34.23449   & & 152-305 \\
GRB200716C & 982707 & 2020-07-16 22:57:41.42 & 86.572 & 15.1708   & & 155-170,175-191 \\
GRB200528A & 974827 & 2020-05-28 10:27:23.84 & 59.984 & 1.923817  & & 60.3-74.5,82.9-200 \\
GRB200306C & 960102 & 2020-03-06 22:50:39.06 & 53.152 & 13.72531  & & 867-1392 \\
GRB200306A & 959917 & 2020-03-06 00:19:40.18 & 32.964 & 1.109263  & & 140-296 \\
GRB200303A & 959431 & 2020-03-03 02:34:57.45 & 94.22 & 6.395724   & & 183-216,218-238 \\
GRB200227A & 958592 & 2020-02-27 07:20:08.00 & 30.328 & 9.202282  & & 264-1281 \\
GRB191031C & 932595 & 2019-10-31 18:43:54.90 & 78.5 & 7.509189    & & 272-459,707-1620 \\
GRB190926A & 926515 & 2019-09-26 09:52:16.35 & 362.34 & 19.60651  & & 527-611 \\
GRB190613B & 908329 & 2019-06-13 10:47:02.05 & 160.82 & 17.08426  & & 213-344 \\
GRB190613A & 908288 & 2019-06-13 04:07:18.31 & 17.632 & 3.113922  & & 179-264 \\
GRB190109A & 882747 & 2019-01-09 05:12:41.78 & 115.008 & 33.8917  & & 189-488 \\
GRB190103B & 881709 & 2019-01-03 21:03:51.18 & 19.52 & 2.716798   & & 128-340 \\
GRB181023A & 868427 & 2018-10-23 02:04:02.20 & 56.984 & 8.033103  & & 405-1330 \\
GRB181013A & 866783 & 2018-10-13 02:39:29.67 & 104.208 & 53.6958  & & 298-2013 \\
GRB181002A & 865036 & 2018-10-02 11:26:43.20 & 63.408 & 12.34832  & & 142-209 \\
GRB180930A & 864584 & 2018-09-30 12:43:49.26 & 6.804 & 0.9557908  & & 306-743 \\
GRB180904A & 859282 & 2018-09-04 21:28:32.41 & 4.632 & 1.031591   & & 129-208,440-691 \\
GRB180821A & 854578 & 2018-08-21 16:40:34.53 & 64.0 & 22.62742    & & 261-452 \\
GRB180818B & 853882 & 2018-08-18 12:30:18.55 & 137.46 & 9.700067  & & 241-370 \\
GRB180809B & 852553 & 2018-08-09 20:28:48.10 & 233.244 & 1.124455 & & 395-408 \\
GRB180706A & 846395 & 2018-07-06 08:24:40.41 & 42.444 & 7.162597  & & 95.1-201 \\
GRB180510A & 831810 & 2018-05-10 19:24:34.30 & 64.0 & 16.0        & & 94.5-113,207-321 \\
GRB180316A & 814677 & 2018-03-16 04:57:25.09 & 87.0 & 20.59126    & & 232-582 \\
GRB180111A & 804692 & 2018-01-11 16:42:06.38 & 50.892 & 1.880136  & & 76.6-197 \\
GRB171120A & 791201 & 2017-11-20 13:20:02.49 & 64.0 & 16.0        & & 229-583 \\
GRB170906A & 770957 & 2017-09-06 00:43:11.62 & 88.112 & 1.444798  & & 169-179 \\
GRB170810A & 767284 & 2017-08-10 22:01:41.06 & 152.42 & 21.49788  & & 315-389,737-993 \\
GRB170714A & 762535 & 2017-07-14 12:25:32.63 & 459.216 & 94.60989 & & 621-638,641-656,658-691,798-831,836-851,852-885,886-928,930-1115,1117-1311,1313-1374,1376-1485 \\
GRB170626A & 758766 & 2017-06-26 09:37:23.12 & 12.948 & 0.3465948 & & 64.8-365,367-382 \\
GRB170318B & 743086 & 2017-03-18 15:27:52.55 & 1.072 & 0.2218468  & & 111-222 \\
GRB170206B & 737125 & 2017-02-06 21:57:25.48 & 13.36 & 2.677025   & & 105-288 \\
GRB170115A & 733103 & 2017-01-15 20:50:08.09 & 32.044 & 9.16922   & & 127-178 \\
GRB170111A & 731887 & 2017-01-11 00:33:28.18 & 29.876 & 9.71117   & & 368-694 \\
GRB161129A & 724438 & 2016-11-29 07:11:39.85 & 35.544 & 2.095133  & & 183-5783 \\
GRB161113A & 722063 & 2016-11-13 17:30:26.74 & 43.464 & 5.282643  & & 61.6-76.1,78.3-102 \\
GRB161022A & 718655 & 2016-10-22 02:43:40.14 & 7.248 & 1.06949    & & 178-1128 \\
GRB161004A & 715084 & 2016-10-04 12:58:28.67 & 1.32 & 0.2997866   & & 173-629 \\
GRB161001A & 714404 & 2016-10-01 01:05:16.68 & 2.6 & 0.3853518    & & 102-214 \\
GRB160912A & 711914 & 2016-09-12 16:07:43.14 & 207.308 & 16.02694 & & 1445-5911 \\
GRB160905A & 711084 & 2016-09-05 11:18:58.38 & 64.0 & 16.0        & & 76.0-110 \\
GRB160506A & 685245 & 2016-05-06 03:29:15.86 & 260.53 & 78.32     & & 339-507 \\
GRB160408A & 682059 & 2016-04-08 06:25:43.77 & 0.32 & 0.04        & & 180-650 \\
GRB160325A & 680436 & 2016-03-25 07:00:03.61 & 61.724 & 12.05326  & & 155-230 \\
GRB160127A & 671828 & 2016-01-27 08:43:07.31 & 6.156 & 0.9366579  & & 82.3-190 \\
GRB160119A & 671014 & 2016-01-19 03:06:08.28 & 120.14 & 14.98463  & & 955-1651 \\
GRB160104A & 669319 & 2016-01-04 11:24:10.84 & 16.564 & 2.787651  & & 110-125,170-446 \\
GRB151228B & 668641 & 2015-12-28 22:47:14.94 & 48.0 & 16.0        & & 103-281 \\
GRB151210A & 666931 & 2015-12-10 03:12:56.46 & 86.96 & 10.67124   & & 267-877 \\
GRB151118A & 664078 & 2015-11-18 03:06:30.01 & 23.568 & 10.67949  & & 94.0-352 \\
GRB151111A & 663074 & 2015-11-11 08:33:23.41 & 76.044 & 12.75371  & & 96.5-357 \\
GRB151001A & 657286 & 2015-10-01 15:04:22.93 & 9.828 & 3.199572   & & 480-1421 \\
GRB150911A & 655262 & 2015-09-11 18:40:21.48 & 7.136 & 0.5997466  & & 92.4-152 \\
GRB150831A & 653838 & 2015-08-31 10:34:12.89 & 0.92 & 0.1162755   & & 75.3-186 \\
GRB150724A & 650141 & 2015-07-24 05:46:35.98 & 280.008 & 43.96185 & & 962-1500 \\
GRB150722A & 649916 & 2015-07-22 09:57:53.11 & 67.312 & 11.76489  & & 4275-11500 \\
GRB150711A & 648601 & 2015-07-11 18:23:03.68 & 70.956 & 14.21344  & & 145-226 \\
GRB150626A & 646603 & 2015-06-26 02:12:49.52 & 97.264 & 34.93727  & & 98.2-199 \\
GRB150616A & 644259 & 2015-06-16 22:49:19.26 & 608.408 & 104.2196 & & 1435-6780 \\
GRB150323C & 636005 & 2015-03-23 17:05:21.49 & 159.664 & 44.61752 & & 176-364,383-515,568-921 \\
GRB150222A & 632180 & 2015-02-22 16:56:40.85 & 15.836 & 6.851308  & & 144-239,250-652 \\
GRB150103A & 623368 & 2015-01-03 20:02:18.99 & 49.088 & 9.807099  & & 140-362 \\
GRB141221A & 622006 & 2014-12-21 08:07:10.95 & 36.816 & 4.103726  & & 170-641 \\
GRB141130A & 620090 & 2014-11-30 23:10:56.07 & 62.856 & 11.37204  & & 76.0-108,161-531 \\
GRB141026A & 616502 & 2014-10-26 02:36:51.29 & 139.476 & 14.83812 & & 160-175 \\
GRB141020A & 615873 & 2014-10-20 07:48:39.28 & 14.896 & 2.165387  & & 319-695 \\
GRB140916A & 612804 & 2014-09-16 10:43:47.46 & 68.976 & 9.45079   & & 181-218 \\
GRB140713A & 604232 & 2014-07-13 18:43:45.11 & 6.016 & 0.728066   & & 577-1568 \\
GRB140710A & 603954 & 2014-07-10 10:16:40.04 & 3.0 & 2.236068     & & 336-4400 \\
GRB140709A & 603810 & 2014-07-09 01:13:41.16 & 105.192 & 7.385089 & & 131-304 \\
GRB140413A & 595616 & 2014-04-13 00:09:40.12 & 132.976 & 14.36711 & & 2320-3100,3144-3344,3516-3772 \\
GRB140108A & 583338 & 2014-01-08 17:18:42.52 & 95.232 & 3.224261  & & 199-383 \\
GRB131127A & 579571 & 2013-11-27 10:11:35.55 & 94.972 & 15.13601  & & 518-5098 \\
GRB131018A & 574935 & 2013-10-18 12:47:48.83 & 73.216 & 18.97341  & & 172-225 \\
GRB130803A & 565263 & 2013-08-03 10:02:52.49 & 43.572 & 1.606636  & & 477-668 \\
GRB130722A & 563213 & 2013-07-22 08:19:19.22 & 98.368 & 12.86089  & & 195-395 \\
GRB130211A & 548276 & 2013-02-11 03:36:32.16 & 30.752 & 6.35498   & & 145-460 \\
GRB130131A & 547407 & 2013-01-31 13:56:22.02 & 51.524 & 2.512194  & & 107-541 \\
GRB121212A & 541371 & 2012-12-12 06:56:13.38 & 6.056 & 1.380598   & & 196-372,512-1011 \\
GRB121125A & 539563 & 2012-11-25 08:32:27.27 & 52.252 & 5.03752   & & 88.7-134 \\
GRB121028A & 536897 & 2012-10-28 05:04:31.91 & 2.876 & 0.3705995  & & 667-1357 \\
GRB120922A & 534394 & 2012-09-22 22:30:28.65 & 168.224 & 29.23869 & & 305-707 \\
GRB120728A & 529021 & 2012-07-28 22:25:11.03 & 20.976 & 2.059219  & & 460-951 \\
GRB120401A & 519043 & 2012-04-01 05:24:15.94 & 130.272 & 24.22921 & & 217-322 \\
GRB120328A & 518792 & 2012-03-28 03:06:19.03 & 29.928 & 6.665064  & & 96.4-160 \\
GRB120308A & 517234 & 2012-03-08 06:13:38.65 & 61.256 & 16.81484  & & 112-305 \\
GRB120102A & 510922 & 2012-01-02 02:15:55.01 & 38.708 & 3.507323  & & 958-6115 \\
GRB111215A & 509717 & 2011-12-15 14:04:08.32 & 373.82 & 93.22025  & & 407-460,576-701,882-1338 \\
GRB111123A & 508319 & 2011-11-23 18:13:21.10 & 290.0 & 88.09086   & & 336-372,427-671 \\
GRB110921A & 503652 & 2011-09-21 13:51:20.16 & 32.548 & 4.442038  & & 189-347,480-758 \\
GRB110915A & 503219 & 2011-09-15 13:20:44.40 & 79.568 & 2.982415  & & 148-231,317-377 \\
GRB110726A & 458059 & 2011-07-26 01:30:40.54 & 5.156 & 1.107307   & & 61.9-121 \\
GRB110709A & 456939 & 2011-07-09 15:24:29.14 & 44.328 & 0.5420701 & & 55.8-74.4,87.1-120 \\
GRB110520A & 453747 & 2011-05-20 20:28:48.19 & 20.864 & 5.367231  & & 185-433 \\
GRB110414A & 451343 & 2011-04-14 07:42:14.52 & 153.116 & 68.9858  & & 282-603 \\
GRB110407A & 450884 & 2011-04-07 14:06:41.16 & 155.568 & 21.16662 & & 402-877 \\
GRB110119A & 442978 & 2011-01-19 22:20:58.37 & 213.928 & 10.21124 & & 370-415,516-5338 \\
GRB101219B & 440635 & 2010-12-19 16:27:53.51 & 41.856 & 5.839452  & & 273-577 \\
GRB100905A & 433442 & 2010-09-05 15:08:14.71 & 3.396 & 0.5036983  & & 228-594 \\
GRB100727A & 430094 & 2010-07-27 05:42:17.78 & 71.9 & 13.01826    & & 136-597 \\
GRB100619A & 424998 & 2010-06-19 00:21:07.24 & 97.7 & 1.624148    & & 870-5397 \\
GRB100526A & 423181 & 2010-05-26 16:26:10.16 & 97.6 & 8.475306    & & 170-256 \\
GRB100413A & 419404 & 2010-04-13 17:33:28.37 & 192.64 & 12.09429  & & 197-230 \\
GRB100212A & 412081 & 2010-02-12 14:07:22.02 & 163.764 & 13.3436  & & 625-844 \\
GRB091221 & 380311 & 2009-12-21 20:52:52.29 & 62.9 & 5.659316     & & 90.6-169 \\
GRB091104 & 374875 & 2009-11-04 08:49:22.76 & 107.136 & 19.64406  & & 186-295 \\
GRB091026 & 373871 & 2009-10-26 13:11:30.15 & 174.0 & 72.52758    & & 619-1030 \\
GRB090904A & 361830 & 2009-09-04 01:01:06.94 & 129.532 & 9.365273 & & 288-451 \\
GRB090831C & 361489 & 2009-08-31 21:30:25.92 & 77.952 & 14.79992  & & 165-235 \\
GRB090807A & 359378 & 2009-08-07 15:00:27.02 & 146.352 & 13.19631 & & 176-258,5577-17500 \\
GRB081102 & 333427 & 2008-11-02 17:44:39.52 & 50.32 & 12.60935    & & 928-1703 \\
GRB080906 & 323984 & 2008-09-06 13:33:16.34 & 148.212 & 19.6103   & & 150-370,564-674 \\
\bottomrule 
\end{longtable}

\begin{center}
\scriptsize
\setlength{\tabcolsep}{7pt} 
\renewcommand{\arraystretch}{1.2} 
\begin{longtable}{l c c c c c c}
\caption{Time-resolved spectral analysis results from the \textit{Fermi} LAT observations during the X-ray flare intervals of the GRB sample. For each flare interval, the table lists the GRB name, the \textit{Fermi} trigger time (Mission Elapsed Time; MET), the flare time interval relative to the \textit{Swift} BAT trigger, the integrated energy flux in the 0.1–10 GeV band (or the corresponding 1 GeV upper limit when no significant detection is obtained), the best-fit photon index ($\Gamma_\gamma$), the Test Statistic (TS), and the corresponding \textit{Fermi} source name if detected by \textit{Fermi}. Fluxes are reported in units of $10^{-9} \mathrm{erg cm^{-2} s^{-1}}$. Spectral parameters are quoted only for intervals with significant LAT detections with flux at 1~GeV in bracket, while intervals with low TS values (or zero) are reported as upper limits.}
\label{tab:lat_all} \\
\toprule
GRB name & Trigger Time & Flare Interval & F$_{0.1-10\,GeV}$      & \(\Gamma_{\gamma}\) & Test Statistic (TS) & Fermi Name \\
    & in Fermi(MET) & (s)           & [erg\,cm$^{-2}$\,s$^{-1}$] & & & \\
    & corresponding to BAT &  & [$\times10^{-9}$] & & & \\
\midrule
\endfirsthead
\toprule
GRB name & Trigger Time & Flare Interval & F$_{0.1-10\,GeV}$      & \(\Gamma_{\gamma}\) & Test Statistic (TS) & Fermi Name \\
    & in Fermi(MET) & (s)           & [erg\,cm$^{-2}$\,s$^{-1}$] & & & \\
    & corresponding to BAT &  & [$\times10^{-9}$] & & & \\
\midrule
\endhead
\midrule
\multicolumn{7}{r}{\textit{Continued on next page}} \\
\endfoot
\endlastfoot
GRB240912A & 747798404.2  & 134-184   & 51.9  &  &  & GRB240912074 \\
           & 	           & 251-772   & 4.68  &  & 1 &  \\
GRB230414B & 703181666.2  & 370-708   & 1.67  &  &  &  \\
GRB210420B & 640636482.1  & 575-989   & 3.87  &  &  &  \\
GRB151027A & 467611108.2  & 262-429   & 5.81  &  &  & GRB151027166 \\
GRB150821A & 461843045.0  & 233-286   & 9.38  &  &  & GRB150821406 \\
           &              &  288-364  & 5.95  &  &  &  \\
           &              & 481-1282  & 0.68  &  &  &  \\
GRB140907A & 431798831.8  & 160-550   & 1.18  &  &  & GRB140907672 \\
GRB131103A & 405209248.8  & 360-600   & 3.46  &  &  &  \\
GRB130925A & 401775087.9  & 740-1179  & 0.79  &  &  & GRB130925164 \\
GRB100728A & 301976306.2  & 194-252   & 15.4  &   & 11  & GRB100728095 \\
           &              & 253-296   & 22.3 +/- 15.8 (4.5 +/- 1.9) & -1.7 +/- 0.4 &  37  &  \\
           &              & 297-359   & 7.45 &    &    &  \\
           &              & 382-421   & 11.1 &      &    &  \\
           &              & 449-487   & 13.2 &     &  5  &  \\
           &              & 514-660   & 7.31 +/- 6.67 (0.24 +/- 0.14) & -0.90 +/- 0.57 &  26  &  \\
           &              & 660-893   & 1.84  &     &    &  \\
GRB080928  & 244306893.9  & 339-405   & 8.57  &  &  &   \\
\hdashline
GRB250516A & 769098933.9 & 207-496  & 2.24   &  & 3 & GRB250516608 \\
GRB250207A & 760583772.3 & 129-176  & 16.4   &  & 8 & GRB250207053 \\
           &             & 180-303  & 6.14   &  & 9 &  \\
GRB230510A & 705413193.6 & 259-313  & 9.86   &  &  & GRB230510503 \\
GRB230328B & 701708093.5 & 94.4-262 & 3.67    &  &  & GRB230328621 \\
GRB220618A & 677236234.9 & 185-245  & 9.81  &  &  & GRB220618382 \\
           &             &  280-403 & 5.55   &  &  &  \\
GRB210724A & 648850454.2 & 170-511  & 3.93   &  &  &  \\
GRB210410A & 639708801.5 & 69-111   & 35.8   &  &  & GRB210410037 \\
GRB201029A & 625695595.1 & 228-910  & 1.53 &  &  & GRB201029847 \\
           &             & 1038-1859& 0.61 &  &  &  \\
GRB200716C & 616633066.4 & 155-170  & 40.9 &  &  & GRB200716957 \\
           &             & 175-191  & 40.3   &  &  &  \\
GRB200528A & 612354448.8 & 83-200   & 4.01   &  &  & GRB200528436 \\
GRB200227A & 604480813.0 & 264-1281 & 0.79   &  &  & GRB200227306 \\
GRB191031C & 594240239.9 & 272-459  & 3.83   &  &  & GRB191031780 \\
GRB190926A & 591184341.4 & 332-445  & 4.21   &  &  &  \\
           &             & 527-611  & 6.06   &  &  &  \\
GRB190613B & 582115627.1 & 87-203   & 6.78  &  &  & GRB190613449 \\
           &             &  213-344 & 7.78   &  &  &  \\
GRB190613A & 582091643.3 & 179-264  & 6.76  &  &  & GRB190613172 \\
GRB190109A & 568703566.8 & 189-563  & 6.96  &  &  & GRB190109217 \\
GRB180706A & 552558285.4 & 95-201   & 6.66   &  &  & GRB180706351 \\
GRB171120A & 532876807.5 & 229-583  & 2.94 +/- 1.11 (0.40 +/- 0.12)  & -2.72 +/- 0.44  & 43  & GRB171120556 \\
GRB170906A & 526351396.6 & 169-178  & 185.0   &    &  3  & GRB170906030 \\
GRB170810A & 524095306.1 & 315-389  & 11.3 &   &    & GRB170810918 \\
           &             & 725-998  & 2.67 +/- 1.52 (0.64 +/- 0.25)  &  -2.4 +/- 0.57  &   15  &  \\
GRB170626A & 520162648.1 & 65-365   & 3.3   &  &  & GRB170626401 \\
           &             & 367-382  & 44.0   &  &  & \\
GRB161001A & 496976720.7 & 102-214  & 5.38   &  &  & GRB161001045 \\
GRB160912A & 495389267.1 & 1445-2300& 0.58   &  &  & GRB160912674 \\
GRB160905A & 494767142.4 & 76-110   & 21.1   &  &  & GRB160905471 \\
GRB160408A & 481789547.8 & 180-650  & 2.21   &  &  & GRB160408268 \\
GRB160325A & 480582007.6 & 155-230  & 6.38 +/- 3.81 (1.38 +/- 0.51)  & -2 +/- 0.00    & 14  & GRB160325291 \\
GRB151111A & 468923607.4 & 96-357   & 4.56   &  &  & GRB151111356 \\
GRB151001A & 465404666.9 & 480-1421 & 1.67   &  &  & GRB151001628 \\
GRB150711A & 458331787.7 & 145-226  & 6.29   &  &  & GRB150711766 \\
GRB150323C & 448823124.5 & 176-364  & 4.02   &  &  & GRB150323712 \\
           &             & 383-515  & 4.59 &  &  &  \\
           &             & 568-781  & 4.35   &  &  &  \\
GRB140713A & 426969828.1 & 577-1568 & 1.21   &  &  & GRB140713780 \\
GRB121125A & 375525150.3 & 89-134   & 13.0   &  &  & GRB121125356 \\
GRB120308A & 352880020.7 & 112-305  & 4.6   &  &  &  \\
GRB110414A & 324459736.5 & 282-603  &  3.1   &  &  &  \\
GRB100619A & 298599669.2 & 870-1800 & 0.25   &  &  & GRB100619015 \\
GRB100212A & 287676444.0 & 625-844  & 4.97    &  &  & GRB100212588 \\
GRB090831C & 273447027.9 & 165-235  & 13.3   &  &  &  \\
\bottomrule  
\end{longtable}
\end{center}

\begin{center}
\scriptsize
\setlength{\tabcolsep}{3pt} 
\renewcommand{\arraystretch}{1.2} 
\begin{longtable}{l c c c c c c c c c}
\caption{Time-resolved spectral analysis results from the \textit{Swift} XRT observations during the X-ray flare intervals of GRBs with simultaneous \textit{Swift} XRT and \textit{Fermi} LAT coverage. For each flare interval, the table lists the observed 0.3--10 keV energy flux ($F_{0.3-10\mathrm{keV}}$), $F_{1\mathrm{keV}}$, $F_{10\mathrm{keV}}$, the X-ray photon index ($\Gamma_{\rm X}$), the fit statistic (C-statistic/degrees of freedom), the intrinsic neutral hydrogen column density ($N_{\rm H}$) in units of $10^{22},\mathrm{cm^{-2}}$, the GRB redshift ($z$), and whether contemporaneous \textit{Swift} BAT observations are available (BAT; Y = yes, N = no). Fluxes are reported in units of $10^{-9} \mathrm{erg,cm^{-2}s^{-1}}$, and uncertainties correspond to the 90\% confidence level.}
\label{tab:xrt_all} \\
\toprule
GRB & Flare Interval & $F_{0.3-10\,\mathrm{keV}}$ & $F_{1\,\mathrm{keV}}$ & $F_{10\,\mathrm{keV}}$ & -$\Gamma_X$ & Fit Statistics & $N_H$ & z & BAT \\
    & (s) & [erg\,cm$^{-2}$\,s$^{-1}$] $\times10^{-9}$ & [erg\,cm$^{-2}$\,s$^{-1}$] $\times10^{-9}$ & [erg\,cm$^{-2}$\,s$^{-1}$] $\times10^{-9}$ & & & [$10^{22}$ cm$^{-2}$] &  & \\
\midrule
\endfirsthead
\toprule
GRB & Flare Interval & $F_{0.3-10\,\mathrm{keV}}$ & $F_{1\,\mathrm{keV}}$ & $F_{10\,\mathrm{keV}}$ &-$\Gamma_X$ & Fit Statistics & $N_H$ & z & BAT \\
    & (s) & [erg\,cm$^{-2}$\,s$^{-1}$] $\times10^{-9}$ & [erg\,cm$^{-2}$\,s$^{-1}$] $\times10^{-9}$ & [erg\,cm$^{-2}$\,s$^{-1}$] $\times10^{-9}$ & & & [$10^{22}$ cm$^{-2}$] &  & \\
\midrule
\endhead
\midrule
\multicolumn{8}{r}{\textit{Continued on next page}} \\
\endfoot
\endlastfoot
GRB240912A & 134-184 & 15.80 +/- 0.50 & 4.5 +/- 0.11 &4.60 +/-  0.26& 1.99 +/- 0.05 & 500/565 & 0.75 +/- 0.13 & 1.23 &  Y \\
           & 251-772 & 2.82 +/- 0.04  & 0.75 +/- 0.01&0.98 +/-  0.03& 1.88 +/- 0.03 & 816/784 & 0.26 +/- 0.40 &      & N \\
GRB230414B & 370-708 & 0.05 +/- 0.01  & 0.02 +/- 0.00&0.01 +/-  0.00& 2.23 +/- 0.22 & 89/109  & < 0.57        & 3.568 & Y \\
GRB210420B & 575-989 & 0.16 +/- 0.02  & 0.03 +/- 0.00&0.09 +/-  0.01& 1.52 +/- 0.16 & 238/254 & < 0.1 & 1.4   & Y \\
GRB151027A & 262-429 & 1.51 +/- 0.06  & 0.48 +/- 0.01&0.26 +/-  0.02& 2.26 +/- 0.06 & 487/529 & 0.49 +/- 0.06 & 0.81 & Y \\
GRB150821A & 233-286 & 4.40 +/- 0.39  & 1.20 +/- 0.08&1.50 +/-  0.21& 1.90 +/- 0.13 & 463/500 & 2.6 +/- 0.4   & 0.755 & Y \\
           & 288-364 & 3.03 +/- 0.15  & 0.76 +/- 0.04&1.20 +/-  0.10& 1.81 +/- 0.09 & 583/578 & 2.0 +/- 0.24  &  & Y \\
           & 481-1282& 1.15 +/- 0.03  & 0.32 +/- 0.01&0.36 +/-  0.02& 1.95 +/- 0.05 & 707/744 & 2.02 +/- 0.12 &  & N \\
GRB140907A & 160-550 & 0.19 +/- 0.02  & 0.04 +/- 0.01&0.09 +/-  0.02& 1.70 +/- 0.24 & 174/190 & < 0.55        & 1.21 & N \\
GRB131103A & 360-600 & 0.15 +/- 0.04  & 0.04 +/- 0.01&0.04 +/-  0.03& 2.0 +/-0.4    & 68/94   & 1.2 +/- 0.7   & 0.599 & Y \\
GRB130925A & 740-1179& 7.02 +/- 0.13  & 1.3 +/- 0.02 &3.30 +/-  0.08& 1.57 +/- 0.03 & 1072/915 & 2.55 +/- 0.08 & 0.347 & Y \\
GRB100728A & 194-252 & 11.10 +/- 0.3  & 2.0 +/- 0.09 &7.20 +/-  0.39& 1.44 +/- 0.07 & 666/718 & 5.2 +/- 0.7   & 1.567 & Y \\
           & 253-296 & 8.2 +/- 0.5    & 2.2 +/- 0.11 &2.70 +/-  0.29& 1.92 +/- 0.10 & 496/573 & 4.9 +/- 0.8   &  & Y \\
           & 297-359 & 12.70 +/- 0.6  & 3.6 +/- 0.13 &3.60 +/-  0.29& 2.00 +/- 0.07 & 598/627 & 5.1 +/- 0.6   &  & Y \\
           & 382-421 & 9.5 +/- 0.7    & 2.9 +/- 0.14 &2.00 +/-  0.25& 2.17 +/- 0.10 & 489/534 & 4.3 +/- 0.7   &  & Y \\
           & 449-487 & 4.8 +/- 0.7    & 1.6 +/- 0.14 &0.48 +/-  0.11& 2.52 +/- 0.14 & 283/374 & 3.6 +/- 0.7   &  & Y \\
           & 514-660 & 7.8 +/- 0.4    & 1.5 +/- 0.04 &0.72 +/-  0.04& 2.45 +/- 0.06 & 720/709 & 3.48 +/- 0.23 &  & Y \\
           & 660-893 & 2.73 +/- 0.12  & 0.85 +/- 0.04&0.52 +/-  0.04& 2.13 +/- 0.12 & 720/709 & 3.48 +/- 0.23 &  & Y \\
GRB080928 & 339-405 & 2.55 +/- 0.09   & 0.56 +/- 0.02&1.30 +/-  0.08& 1.64 +/- 0.07 & 511/572 & 0.82 +/- 0.24 & 1.69 & Y \\
\hdashline
GRB250516A & 207-496 & 0.90 +/- 0.05  & 0.29 +/- 0.01&0.12 +/-  0.02& 2.41 +/- 0.13 & 388/418 & 0.28 +/- 0.05    & & N \\  
GRB250207A & 129-176 & 1.19 +/- 0.07  & 0.31 +/- 0.01&0.44 +/-  0.05& 1.85 +/- 0.12 & 332/387  &  0.02 +/- 0.03 & & Y  \\
           & 180-303 & 1.35 +/- 0.06  & 0.43 +/- 0.01&0.23 +/-  0.02& 2.27 +/- 0.08 & 409/481 & 0.09 +/- 0.02   & & Y  \\
GRB230510A & 259-313 & 0.70 +/- 0.06  & 0.2 +/- 0.01 &0.20 +/-  0.03& 2.01 +/- 0.12 & 334/340 & < 0.16          & & N \\  
GRB230328B & 94-262  & 0.48 +/- 0.06  & 0.13 +/- 0.02&0.16 +/-  0.05& 1.92 +/- 0.26 & 141/179 & 0.05 +/- 0.07    & & Y  \\
GRB220618A & 185-245 & 1.58 +/- 0.24  & 0.46 +/- 0.05&0.39 +/-  0.10& 2.08 +/- 0.21 & 316/341 & 0.07 +/- 0.13   & & N \\
           & 280-403 & 4.46 +/- 1.19  & 1.3 +/- 0.22 &0.08 +/-  0.03& 3.22 +/- 0.18 & 331/336 & 0.29 +/- 0.09   & & N \\
GRB210724A & 170-511 & 1.11 +/- 0.04  & 0.36 +/- 0.01&0.14 +/-  0.02& 2.40 +/- 0.08 &  484/493 & 0.14 +/- 0.25  & & Y  \\ 
GRB210410A & 69-111  & 1.99 +/- 0.14  & 0.27 +/- 0.03&1.70 +/-  0.15& 1.21 +/- 0.13 & 355/444 & 0.11 +/- 0.6     & & Y  \\
GRB201029A & 228-910 & 4.28 +/- 0.09  & 0.94 +/- 0.03&2.10 +/-  0.09& 1.64 +/- 0.05 & 740/762 & 0.5 +/- 0.03    & & Y  \\ 
           &1038-1859& 0.49 +/- 0.05  & 0.09 +/- 0.01&0.30 +/-  0.05& 1.50 +/- 0.21 & 207/266 & 0.29 +/- 0.12& & N \\
GRB200716C & 155-170 & 3.95 +/- 0.26  & 0.80 +/- 0.06&2.20 +/-  0.25& 1.56 +/- 0.13 & 402/407 & 0.01 +/- 0.01  & & Y  \\
           & 175-191 & 3.13 +/- 0.23  & 0.75 +/- 0.06&1.40 +/-  0.20& 1.74 +/- 0.15 & 326/377 & 0.12 +/- 0.05  & & Y  \\
GRB200528A & 83-200  & 9.1 +/- 0.4    & 2.9 +/- 0.08 &1.40 +/-  0.11& 2.33 +/- 0.06 & 679/641 & 0.42 +/- 0.26     & & Y  \\
GRB200227A & 264-1281& 0.24 +/- 0.03  & 0.06 +/- 0.01&0.10 +/-  0.02& 1.79 +/- 0.20 & 282/315 & 0.36 +/- 0.14   & & N \\
GRB191031C & 272-459 & 38.6 +/- 0.02  & 10.0 +/- 0.54&14.00 +/-  1.90& 1.86 +/- 0.13 & 339/432 &  0.01 +/- 0.01   & & Y  \\
GRB190926A & 332-445 & 6.22 +/- 0.15  & 1.5 +/- 0.04 &2.70 +/-  0.13& 1.73 +/- 0.05 & 667/706 & 0.32 +/- 0.03  & & Y  \\
           & 527-611 & 2.89 +/- 0.13  & 0.86 +/- 0.03&0.71 +/-  0.07&  2.08 +/- 0.08 & 534/558 & 0.27 +/- 0.03 & & Y  \\
GRB190613B & 87-203  & 8.62 +/- 0.26  & 2.3 +/- 0.06 &2.80 +/-  0.18& 1.92 +/- 0.06 & 591/657 & 0.37 +/- 0.03   & & Y  \\
           & 213-344 & 1.33 +/- 0.07  & 0.39 +/- 0.01&0.34 +/-  0.04& 2.06 +/- 0.11 & 464/481 &  0.21 +/- 0.04 & & Y  \\
GRB190613A & 179-264 & 39.2 +/- 0.03  & 13.0 +/- 0.16&5.20 +/-  1.40& 2.39 +/- 0.19 & 260/280 &  0.09 +/- 0.04 & & N \\
GRB190109A & 189-563 & 1.50 +/- 0.03  & 0.41 +/- 0.01&0.47 +/-  0.02& 1.94 +/- 0.05 & 633/640 & 0.07 +/- 0.01   & & Y  \\
GRB180706A & 95-201  & 2.66 +/- 0.20  & 0.81 +/- 0.03&0.08 +/-  0.00& 3.03 +/- 0.03 & 476/351 & 0.17 +/- 0.19   & & N \\
GRB171120A & 229-583 & 1.40 +/- 0.68  & 0.28 +/- 0.13&0.71 +/-  0.42& 1.6 +/- 0.5 & 161/189 & 3.13 +/- 1.07    & & Y  \\
GRB170906A & 169-178 & 5.2 +/- 0.9    & 1.6 +/- 0.18 &0.90 +/-  0.27& 2.26 +/- 0.21 & 233/267 & 0.31 +/- 0.10     & & Y  \\
GRB170810A & 315-389 & 0.52 +/- 0.05  & 0.17 +/- 0.01&0.03 +/-  0.01& 2.78 +/- 0.21 & 222/251 & 0.08 +/- 0.04    & & N \\
           & 725-998 & 0.16 +/- 0.03  & 0.05 +/- 0.01&0.03 +/-  0.01& 2.19 +/- 0.17 & 156/135 & < 0.02           & & N \\
GRB170626A & 65-365  & 3.26 +/- 0.06  & 0.99 +/- 0.01&0.73 +/-  0.03& 2.13 +/- 0.03 & 771/688 & 0.06 +/- 0.07   & & N \\
           & 367-382 & 1.09 +/- 0.02  & 0.33 +/- 0.02&0.25 +/-  0.07& 2.13 +/- 0.23 & 175/221 & <0.03           & & N \\
GRB161001A & 102-214 & 0.69 +/- 0.06  & 0.21 +/- 0.01&0.15 +/-  0.04& 2.14 +/- 0.19 & 314/364 & 0.5 +/- 0.1     & & Y  \\
GRB160912A &1445-2300& 0.12 +/- 0.01  & 0.03 +/- 0.00&0.05 +/-  0.01& 1.70 +/- 0.20 & 219/290 & 0.12 +/- 0.11  & & N \\
GRB160905A & 76-110  & 10.7 +/- 0.5   & 2.4 +/- 0.13 &5.20 +/-  0.43& 1.66 +/- 0.09 & 607/649 & 0.23 +/- 0.07   & & Y  \\
GRB160408A & 180-650 & 0.02 +/- 0.01  & 0.01 +/- 0.00&0.01 +/-  0.00& 2.0 +/- 0.5 & 72/77 & 0.05 +/- 0.12       & & N \\
GRB160325A & 155-230 & 2.46 +/- 0.27  & 0.8 +/- 0.05 &0.34 +/-  0.06& 2.37 +/- 0.12 & 469/470 & 0.01 +/- 0.07   & & Y  \\
GRB151111A & 96-357  & 1.13 +/- 0.03  & 0.2 +/- 0.00 &0.76 +/-  0.03& 1.41 +/- 0.06 & 615/673 & 0.08 +/- 0.02   & & Y  \\
GRB151001A & 480-1421& 0.07 +/- 0.01  & 0.02 +/- 0.00&0.04 +/-  0.01& 1.61 +/- 0.18 & 240/249 & 0.03 +/- 0.05  & & N \\
GRB150711A & 145-226 & 1.54 +/- 0.21  & 0.49 +/- 0.04&0.07 +/-  0.02& 2.85 +/- 0.19 & 346/349 & 0.43 +/- 0.07  & & Y  \\
GRB150323C & 176-364 & 2.79 +/- 0.08  & 0.9 +/- 0.02 &0.41 +/-  0.03& 2.34 +/- 0.05 & 605/607 & 0.18 +/- 0.14  & & Y  \\
           & 383-515 & 1.00 +/- 0.31  & 0.13 +/- 0.04&0.01 +/-  0.00& 4.4 +/- 0.4 & 132/160 & 0.26 +/- 0.07    & & N \\
           & 568-781 & 0.09 +/- 0.04  & 0.02 +/- 0.01&0.01 +/-  0.00& 3.9 +/- 0.9 & 64/65 & 0.15 +/- 0.13       & & N \\
GRB140713A & 577-1568& 0.26 +/- 0.02  & 0.05 +/- 0.01&0.15 +/-  0.02& 1.56 +/- 0.18 & 292/346 &  0.49 +/- 0.14  & & N \\ 
GRB121125A & 89-134  & 1.69 +/- 0.09  & 0.45 +/- 0.02&0.58 +/-  0.07& 1.89 +/- 0.12 & 368/444 & 0.13 +/- 0.04  & & Y  \\
GRB120308A & 112-305 & 2.04 +/- 0.35  & 0.65 +/- 0.06&0.08 +/-  0.01& 2.91 +/- 0.08 & 429/443 & 0.12 +/- 0.15  & & N \\
GRB110414A & 282-603 & 0.18 +/- 0.03  & 0.05 +/- 0.01&0.06 +/-  0.01& 1.92 +/- 0.23 & 140/182 & < 0.08         & & N \\
GRB100619A & 870-1800& 2.43 +/- 0.15  & 0.72 +/- 0.03&0.60 +/-  0.04& 2.08 +/- 0.05 & 636/733 &  0.7 +/- 0.4   & & N \\
GRB100212A & 625-844 & 0.17 +/- 0.05  & 0.05 +/- 0.01&0.01 +/-  0.00& 2.91 +/- 0.34 & 119/118 & < 0.07          & & N \\
GRB090831C & 165-235 & 0.46 +/- 0.24  & 0.12 +/- 0.05&0.12 +/-  0.13& 2.0 +/- 0.7 & 48/50 & 0.08 +/- 0.47       & & Y \\ 
\bottomrule  
\end{longtable}
\end{center}

\begin{center}
\scriptsize
\setlength{\tabcolsep}{24.5pt} 
\renewcommand{\arraystretch}{1.2} 
\begin{longtable}{l c c c c }
\caption{Time-resolved spectral analysis results from the \textit{Swift} BAT observations during the X-ray flare intervals of GRBs with contemporaneous BAT coverage and an XRT flux exceeding $1\times10^{-9},\mathrm{erg,cm^{-2},s^{-1}}$. For each flare interval, the table lists the observed 15--150 keV energy flux ($F_{15-150,\mathrm{keV}}$), the BAT photon index ($\Gamma_{\rm B}$), and the fit statistic ($\chi^2$/degrees of freedom). Fluxes are reported in units of $10^{-9} \mathrm{ergcm^{-2}s^{-1}}$, and uncertainties correspond to the 90\% confidence level. For intervals without a statistically significant BAT detection, 15--150 keV flux upper limits (U.L.) are reported in place of spectral parameters.}
\label{tab:batresult} \\
\toprule
GRB & Flare Interval & $F_{15-150\,\mathrm{keV}}$ & -$\Gamma_B$ & Fit Statistics \\
    & (s) & [erg\,cm$^{-2}$\,s$^{-1}$] $\times10^{-9}$ & &\\
\midrule
\endfirsthead
\toprule
GRB & Flare Interval & $F_{15-150\,\mathrm{keV}}$                 & -$\Gamma_B$ & Statistics \\
    & (s)            & [erg\,cm$^{-2}$\,s$^{-1}$] $\times10^{-9}$ & &  \\
\midrule
\endhead
\midrule
\multicolumn{5}{r}{\textit{Continued on next page}} \\
\endfoot
\endlastfoot
GRB240912A & 134-184 & 6.10 +/- 0.85 & 2.29 +/- 0.24 & 78/56\\
GRB150821A & 233-286 & 2.74 (U.L.) &  & 48/57\\
           & 288-364 & 2.18 (U.L.) &  & 43/57\\
GRB130925A & 740-1179 & 6.52 +/- 0.25 & 1.84 +/- 0.07 & 69/56\\
GRB100728A & 194-252 & 11.2 +/- 0.74 & 1.96 +/- 0.11 & 72/56\\ 
           & 253-296 & 4.16+/-0.77 & 2.31+/- 0.33 & 45/56 \\ 
           & 514-660 & 0.87+/-0.30 & 1.62+/-0.55 & 46/56 \\
GRB08928A  & 339-405 & 1.56 (U.L.) &  & 58/57\\          
\hdashline
GRB210410A & 69-111 & 3.37 +/- 1.05 & 1.92 +/- 0.51 & 65/56\\
GRB201029A & 228-910 & 1.56 +/- 0.23 & 1.79 +/- 0.24 & 47/56\\
GRB200716C & 155-170 & 1.94 (U.L.) &  & 63/57\\
           & 175-191 & 2.88 (U.L.) &  & 65/57\\
GRB200528A & 83-200 & 3.92 +/- 0.55 & 2.34 +/- 0.25 & 61/56\\
GRB191031C & 272-459 & 0.65 (U.L.) &  & 75/57\\
GRB190926A & 332-445 & 4.73 +/- 0.54 & 2.36 +/- 0.21 & 39/56\\
GRB190613B & 87-203 & 5.33 +/- 0.80 & 2.03 +/- 0.29 & 66/56\\
GRB190613A & 229-583 & 2.01 (U.L.) &  & 51/57\\ 
GRB171120A & 229-583 & 0.58 (U.L.) &  & 60/57\\ 
GRB160905A & 76-110 & 11.94 +/- 1.21 & 1.76 +/- 0.17 & 52/56\\
GRB160325A & 155-230 & 0.64 (U.L.) &  & 49/57\\

\hline  
\end{longtable}
\end{center}

\begin{table}[h]
\centering
\small
\setlength{\tabcolsep}{0.4pt}
\renewcommand{\arraystretch}{1.25}
\caption{Comparison of the synchrotron self-Compton (SSC) microphysical parameters ($p$, $\epsilon_{\rm e}$ $\epsilon_{\rm B}$, Medium, and $A_*$) inferred from TeV-detected GRBs with the benchmark parameter set adopted in this work for external forward-shock afterglow modeling. The comparison highlights the range of microphysical parameters used in previous TeV afterglow studies and the values adopted here for comparing with flare emission.}
\label{tab:ssc_params_comparison}
\begin{tabular}{l c c c c p{2.5cm} l}
\hline
GRB  & $p$ & $\epsilon_e$ & $\epsilon_B$ & Medium & $A_{*}$ & Reference\\
\hline

190114C  & 2.6 & $7\times10^{-2}$ & $8\times10^{-5}$ & Homogeneous & 0.5 & \citet{MAGIC:2019irs} \\

190114C  & $2.63^{+0.05}_{-0.05}$ & $4.73^{+0.94}_{-1.22}\times10^{-2}$ &
$3.65^{+0.18}_{-0.33}\times10^{-4}$  & Wind & $5.52^{+1.25}_{-0.82}\times10^{-3}$ & \citet{Aguilar-Ruiz:2025ect}\\ 

190114C & $2.5$ & $\approx 0.1$ & $\approx 6 \times 10^{-3} $  & Wind $\&$ Homogeneous &$\approx 0.1$ & \citet{DP2021}\\ 

190829A   & 2.01 & $3^{+2.9}_{-1.7}\times10^{-2}$ & $2.5^{+3.5}_{-1.3}\times10^{-5}$ & Homogeneous & & \citet{2022ApJ...931L..19S}\\

201216C   & $2.1^{+0.5}_{-0.05}$ & $0.08^{+0.82}_{-0.07}$ &$2.5\times10^{-3}$  & Wind & $2.5\times10^{-2}$ & \citet{Abe:2023nhj}\\

221009A   &$2.7^{+0.1}_{-0.2}$ & $\lesssim1.3\times10^{-2}$ & $\lesssim1.6\times10^{-5}$ & Homogeneous & &\citet{Banerjee:2024hxp}\\

221009A   &&&$\sim3\times10^{-3}$ & Wind && \citet{Khangulyan:2023srq}\\

GRB population   &2.2 &0.1 &$10^{-4}$  & Wind & 0.1 &\citet{2025arXiv251005239T}\\
\hline

\end{tabular}
\end{table}

\end{document}